\pdfoutput=1

\documentclass[11pt,twoside,a4paper,cmspaper,final,collab]{cms-tdr}
\def\svnVersion{231011}\def\cmsCernNo{2013-037}\def\cmsCernDate{\today}\def\cmsMessage{Published in the Journal of High Energy Physics as \href{http://dx.doi.org/10.1007/JHEP07(2013)178}{\doi{10.1007/JHEP07(2013)178.}}}

\begin{document}\cmsNoteHeader{EXO-12-009}

\hyphenation{had-ron-i-za-tion}
\hyphenation{cal-or-i-me-ter}
\hyphenation{de-vices}

\RCS$Revision: 192230 $
\RCS$HeadURL: svn+ssh://alverson@svn.cern.ch/reps/tdr2/papers/EXO-12-009/trunk/EXO-12-009.tex $
\RCS$Id: EXO-12-009.tex 192230 2013-06-24 19:06:28Z aferapon $
\newlength\cmsFigWidth
\ifthenelse{\boolean{cms@external}}{\setlength\cmsFigWidth{0.85\columnwidth}}{\setlength\cmsFigWidth{0.4\textwidth}}
\ifthenelse{\boolean{cms@external}}{\providecommand{\cmsLeft}{top}}{\providecommand{\cmsLeft}{left}}
\ifthenelse{\boolean{cms@external}}{\providecommand{\cmsRight}{bottom}}{\providecommand{\cmsRight}{right}}

\cmsNoteHeader{EXO-12-009} 
\title{Search for microscopic black holes in pp collisions at $\sqrt{s} = 8$\TeV}

\providecommand{\PZ}{\ensuremath{\mathrm{Z}}}
\providecommand{\PV}{\ensuremath{{V}}}
\providecommand{\MPL}{\ensuremath{M_\mathrm{Pl}}\xspace}
\providecommand{\MPLtwo}{\ensuremath{M_\mathrm{Pl}^2}\xspace}
\providecommand{\MDtwo}{\ensuremath{M_\mathrm{D}^{n+2}}\xspace}
\providecommand{\MDmin}{\ensuremath{M_\mathrm{D}^\text{min}}\xspace}
\providecommand{\ST}{\ensuremath{S_\mathrm{T}}\xspace}
\providecommand{\TH}{\ensuremath{T_\mathrm{H}}\xspace}
\providecommand{\CLs}{\ensuremath{\mathrm{CL}_\mathrm{s}}\xspace}
\providecommand{\MBH}{\ensuremath{M_\mathrm{BH}}\xspace}     
\providecommand{\MBHmin}{\ensuremath{M_\mathrm{BH}^\text{min}}\xspace}     

\date{\today}

\abstract{
A search for microscopic black holes and string balls is presented, based on a data sample of pp collisions at $\sqrt{s} = 8$\TeV recorded by the CMS experiment at the Large Hadron Collider and corresponding to an integrated luminosity of 12\fbinv. No excess of events with energetic multiparticle final states, typical of black hole production or of similar new physics processes, is observed. Given the agreement of the observations with the expected standard model background, which is dominated by QCD multijet production, 95\% confidence level limits are set on the production of semiclassical or quantum black holes, or of string balls, corresponding to the exclusions of masses below 4.3 to 6.2\TeV, depending on model assumptions. In addition, model-independent limits are set on new physics processes resulting in energetic multiparticle final states.
}

\hypersetup{%
pdfauthor={CMS Collaboration},%
pdftitle={Search for microscopic black holes in pp collisions at sqrt(s) = 8 TeV},%
pdfsubject={CMS},%
pdfkeywords={CMS, physics, black holes, extra dimensions}}

\maketitle 

\section{Introduction\label{s:intro}}
Theoretical models with low-scale quantum gravity aim to account for the origin of the large difference between the electroweak scale ($\sim$0.1\TeV)  and the Planck scale ($\MPL\sim10^{16}$\TeV), known as the hierarchy problem of the standard model (SM) of particle physics. One of the predictions of such scenarios is the possibility of producing microscopic black holes or their quantum precursors in proton-proton collisions at the CERN Large Hadron Collider (LHC)~\cite{dl,gt}.

The basis of this analysis is the theoretical model proposed by Arkani-Hamed, Dimopoulos, and Dvali (ADD)~\cite{add,add1}. This model attempts to solve the hierarchy problem by introducing $n$ large, flat, extra spatial dimensions, compactified on an $n$-dimensional torus or a sphere. By opening the multidimensional space only to the gravitational interaction, the fundamental Planck scale in $4+n$ dimensions, $\MD$, is lowered to the electroweak symmetry breaking scale, such that $\MDtwo \propto \MPLtwo R^{-n}$ where $R$ is the radius of the extra dimensions. The reduction in $\MD$ is accomplished without affecting tight constraints coming from precision measurements of properties of other types of fundamental interactions. The enhanced gravity in multidimensional space allows the formation of microscopic black holes. Production of black holes is also possible in the Randall-Sundrum (RS) model~\cite{rs1,rs2,mr} and in models with unparticles~\cite{Georgi,Mureika}. In the former case, the hierarchy problem is addressed by adding a single compact spatial dimension with radius comparable to the Planck length and with the metric of the 5-dimensional space-time being exponentially ``warped'' in the direction of this extra dimension (the anti-deSitter space-time).

We present an extension of previous searches for microscopic semiclassical black holes~\cite{dl,gt}, quantum black holes~\cite{mr,Calmet}, and string balls~\cite{SB} conducted by the Compact Muon Solenoid (CMS) Collaboration at the CERN LHC~\cite{CMSBH,bh_full2011_jhep}. The analysis utilizes a data sample corresponding to an integrated luminosity of $12.1 \pm 0.5$\fbinv, collected by the CMS detector in pp collisions at a centre-of-mass energy of 8\TeV. These data were recorded in the 2012 running period of the LHC. The 14\% increase in the energy of the machine relative to the 2011 dataset would result in a larger production cross section for black holes and other new physics with energetic multiparticle final states. This allows the current search to penetrate a previously unexplored regime. Searches for black holes have also been performed by the CMS collaboration in the dijet channel~\cite{dijet2011} and by the ATLAS Collaboration~\cite{ATLASbh,ATLASdijet,ATLASlj}.

A characteristic signature of evaporating semiclassical black holes or string balls is a large number of energetic final-state particles of various types, while quantum black holes typically decay into a few energetic partons. Further details of the analysis method and the underlying models can be found in earlier publications~\cite{CMSBH,bh_full2011_jhep}. The results are presented in terms of a set of benchmark scenarios and are also interpreted in terms of model-independent limits on the production cross section for a new physics phenomenon multiplied by the branching fraction of its decay into a multiparticle final state.

\section{The CMS detector\label{s:cmsdetector}}

A detailed description of the CMS detector can be found elsewhere~\cite{CMS}. The central feature of CMS is a 3.8\unit{T} superconducting solenoid of 6\unit{m} internal diameter that encloses a silicon pixel and strip tracker, a lead-tungstate crystal electromagnetic calorimeter (ECAL), and a brass/scintillator hadron calorimeter (HCAL). Muons are detected in gas-ionisation detectors embedded in the steel flux return yoke of the magnet.

The ECAL is a finely segmented calorimeter that uses crystals situated in a barrel region ($\abs{\eta} < 1.48 $) and two endcaps that extend to $\abs{\eta} = 3.0$. Here pseudorapidity $\eta$ is defined as $-\ln[\tan(\theta/2)]$, where $\theta$ is the polar angle measured from the geometrical centre of the detector with respect to the anticlockwise proton beam. The transverse dimensions of the lead-tungstate crystals are $\Delta \eta \times \Delta \phi = 0.0174\times0.0174$, where $\phi$ is the azimuthal angle in radians. The HCAL consists of interleaved brass plates and scintillator sheets that extend to $\abs{\eta} = 3.0 $. The granularity of the HCAL towers is $\Delta \eta \times \Delta \phi = 0.087\times0.087$. Extensive forward calorimetry complements the coverage provided by the barrel and endcap detectors.

The CMS trigger system is composed of two levels that are used to select potentially interesting events. The first level (L1) trigger ensures negligible dead time and is responsible for reducing the event rate to 100\unit{kHz} using the information from calorimeters and muon detectors. After L1 triggering, the data are passed to the software-based high-level trigger, which decreases the event rate to several hundred Hz for further storage.

The data used for this analysis are collected with a set of triggers based on the scalar sum of the transverse energies ($\HT$) of the calorimeter jets found by the trigger. The thresholds for the $\HT$ triggers increased from 200 to 750\GeV, depending on the data-taking period, to cope with the increasing instantaneous luminosity of the LHC. For the earlier part of the data taking, we additionally utilized $\HT$ triggers that use jets reconstructed using the particle-flow (PF) technique~\cite{pf}, which are corrected for the calorimeter response to calculate the $\HT$ variable. The trigger is measured to be fully efficient for jet-enriched collision events with $\HT$ above 1\TeV.

\section{Event reconstruction and Monte Carlo samples\label{s:eventreco}}

The PF technique is used offline to reconstruct and identify charged and neutral particles using information from all the subdetectors. Jets are reconstructed by clustering the PF candidates using an infrared-safe anti-\kt algorithm with a distance parameter of 0.5~\cite{anti-kt,fastjet,fastjet_code}. In the presence of multiple interactions per beam crossing (``pileup"), we identify the primary vertex in the event as the one that has the highest $\sum \pt^2$ of tracks associated with it. Only charged particles originating from the chosen primary vertex are clustered in the jets. The estimated contribution from the neutral-particle energy from the pileup interactions is subtracted on event-by-event basis~\cite{JES}, using \textsc{FastJet} algorithm~\cite{fastjet}, making the analysis insensitive to the effects of the pileup. Additional selection criteria are applied to jets to remove noise and non-collision background~\cite{JES}. The PF jets are required to have transverse momentum $\pt > 50$\GeV and to lie within $\abs{\eta} <  2.6$. The jet energy response is further corrected using simulated events, as well as dijet and photon+jet collision events~\cite{JES}.

Muons are reconstructed using the PF algorithm by matching the tracks in the silicon detector to segments in the muon chambers. Muons with $\abs{\eta} < 2.1$ and $\pt > 50$\GeV are selected. Furthermore, they are required to have an impact parameter less than 0.2 cm to suppress the cosmic ray muon background. In addition, the scalar sum of charged and neutral particle transverse energies, calculated in a cone of $\Delta R = \sqrt{(\Delta \phi)^2 + (\Delta \eta)^2} = 0.3$ around the muon direction, should not exceed 20\% of the muon transverse momentum.

Electrons and photons depositing energy in the ECAL are identified via clustering algorithms, taking into account the expected cluster shapes. Electron reconstruction uses the PF algorithm~\cite{pf,pflep} and requires a silicon tracker trajectory to match an energy cluster found in the ECAL. Photon reconstruction uses ECAL clusters and requires no matching hits in the pixel tracker and an ECAL deposit with a shape consistent with that expected for a photon. Both objects are required to have $\pt > 50$\GeV and to lie within the fiducial region of the barrel ($\abs{\eta} < 1.44$) or endcap ($1.56 < \abs{\eta} < 2.4$). The barrel-endcap transition region is excluded because the reconstruction of electrons and photons in this region is not optimal; energetic electrons and photons in this region nevertheless contribute to the reconstruction of jets. The separation $\Delta R$ between the electron candidate and any muon candidate that has more than 10 hits in the inner tracker is required to be greater than 0.1. We also require the scalar sum of charged and neutral particle transverse energies, calculated in a cone of $\Delta R = 0.3$ around the electron direction, not to exceed 20\% of the electron transverse momentum. The ratio of HCAL to ECAL energy deposits is required to be less than 5\% for photon candidates. Photons must be isolated in the tracker, ECAL, and HCAL. The scalar sums of transverse energy (momenta in the case of the tracker) are calculated in a cone of $\Delta R = 0.4$ around the candidate photon direction. These sums should not exceed 2.0, 4.2, and 2.2\GeV for the tracker, ECAL, and HCAL, respectively.

The missing transverse energy (\MET) is defined as the absolute value of the vector sum of transverse momenta of all the PF objects reconstructed in an event. The \MET~measurement is corrected to account for the jet energy scale calibration~\cite{MET}.

The minimum separation between any two objects (jet, lepton, or photon) in the event is required to be $\Delta R > 0.3$.

Simulated samples of semiclassical black hole events are produced using the parton-level \textsc{BlackMax} v2.01 \cite{BlackMax,BlackMax1} and \textsc{Charybdis} v1.0.3~\cite{CHARYBDIS,CHARYBDIS2} Monte Carlo event generators. Various models are simulated, including black holes that are produced nonrotating or rotating; those with or without mass and angular momentum loss at the time of formation; and those with or without a stable or ``boiling" (i.e., evaporating at a fixed Hawking temperature) remnant. In addition, the modified \textsc{BlackMax} generator settings~\cite{Gingrich} are used to simulate the production of string balls. Detailed descriptions of each model can be found in Refs.~\cite{CMSBH,bh_full2011_jhep}. All these signal samples are generated using the MSTW2008lo68cl~\cite{MSTW} parton distribution functions (PDF). Samples of quantum black holes are generated with the \textsc{qbh}~v1.03 parton-level generator~\cite{qbh_gen,qbh1} using the CTEQ6L PDF set~\cite{CTEQ}. The parton-level events produced by these generators are then used as input to the \PYTHIA v6.426~\cite{PYTHIA} parton showering simulation and a fast parametric simulation of the CMS detector~\cite{FastSim,FastSim1}. The fast simulation was validated with the full detector simulation, based on the  \GEANTfour \cite{GEANT4} framework, for several benchmark points.

The small backgrounds from $\gamma+\text{jets}$, $\PW/\PZ+\text{jets}$ (collectively referred to as $\PV+\text{jets}$ in what follows), and \ttbar production are estimated from Monte Carlo simulations using the \MADGRAPH v5 \cite{madgraph} matrix element event generator interfaced with the \PYTHIA parton showering simulation, followed by the full detector simulation using \GEANTfour. These background samples are generated using the CTEQ6L PDF set. Other SM backgrounds are negligible and therefore were not accounted for in the analysis.

\section{Analysis method\label{s:analysismethod}}

The analysis strategy is identical to the one used in the previous analysis at $\sqrt{s} = 7$\TeV and is described in detail in Ref. \cite{bh_full2011_jhep}. The search for black holes and string balls is performed using events with at least two jets and any number of photons and leptons. There is no explicit requirement of missing transverse energy in the event.

The search for black holes is based on a search for a deviation from the SM background predictions in the $\ST$ spectra observed in data. The $\ST$ variable is defined as the scalar sum of transverse energies of all the final-state objects in the event (jets, leptons, and photons) in excess of 50\GeV. If the \MET\ in the event exceeds 50\GeV, its value is also added to the $\ST$ variable. We then determine the multiplicity $N$ of the objects in the final state by counting all the objects in the events (excluding \MET) that enter the calculation of $\ST$~\cite{CMSBH,bh_full2011_jhep}. We analyse the data for various inclusive multiplicity bins, from $N \ge 2$ to $N \ge 10$, and look for deviations from the SM background predictions in each of these bins.

We use object definitions and isolation requirements for leptons and photons as described in Section~\ref{s:eventreco}. While the isolation requirements are not explicitly used in the analysis (as a non-isolated photon or lepton will be reconstructed as a jet and therefore does not change the values of $\ST$ or $N$ in the event), we keep this approach and disambiguate isolated leptons and photons from jets in order to allow clearer interpretation of a signal should one be observed. Indeed, if an excess in the data were observed, the relative fractions of prompt leptons, photons, and jets in the events responsible for the excess could shed light on the nature of the observed signal.

\begin{figure}[htbp]
\centering
\includegraphics[width=0.49\textwidth]{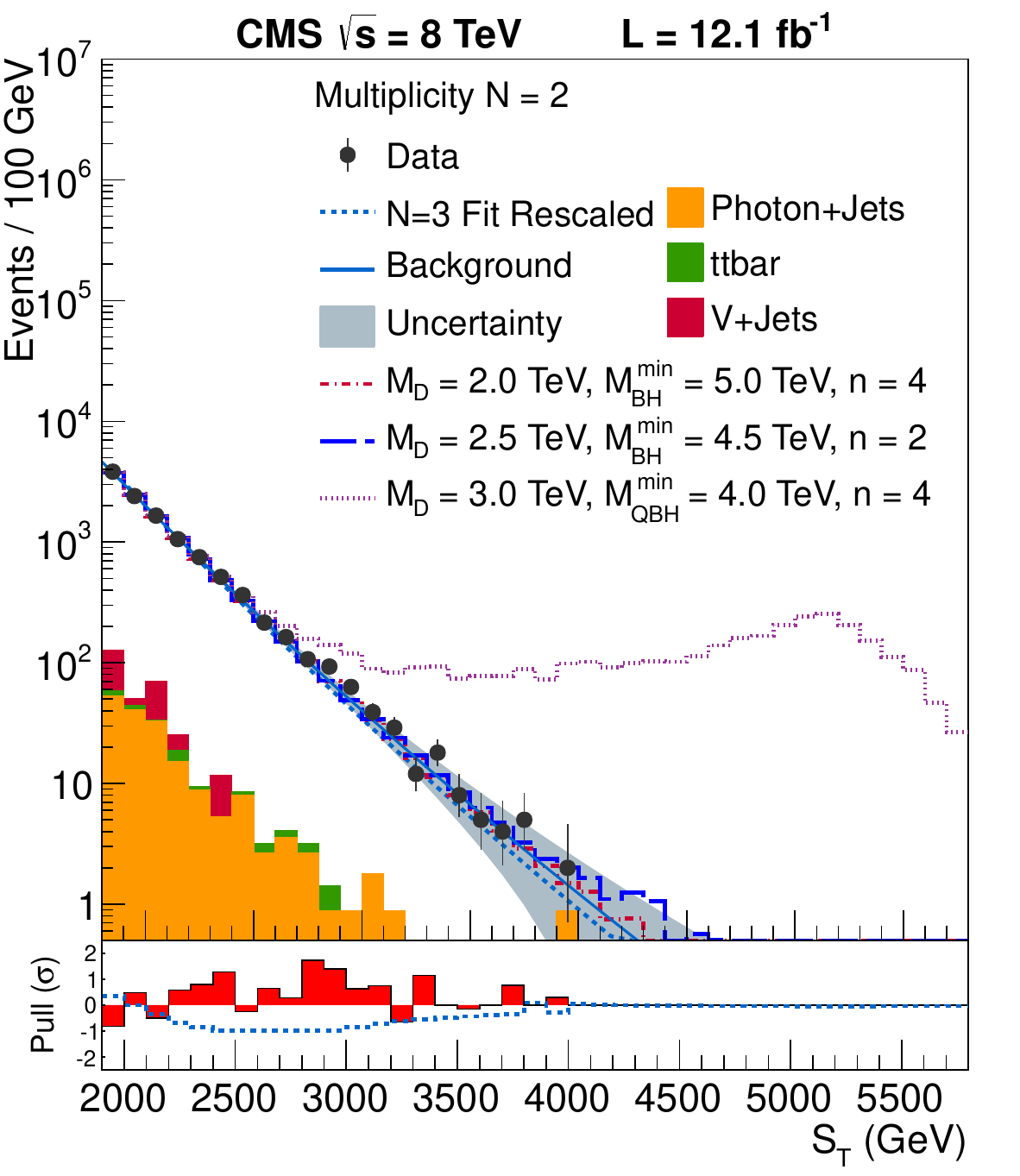}
\includegraphics[width=0.49\textwidth]{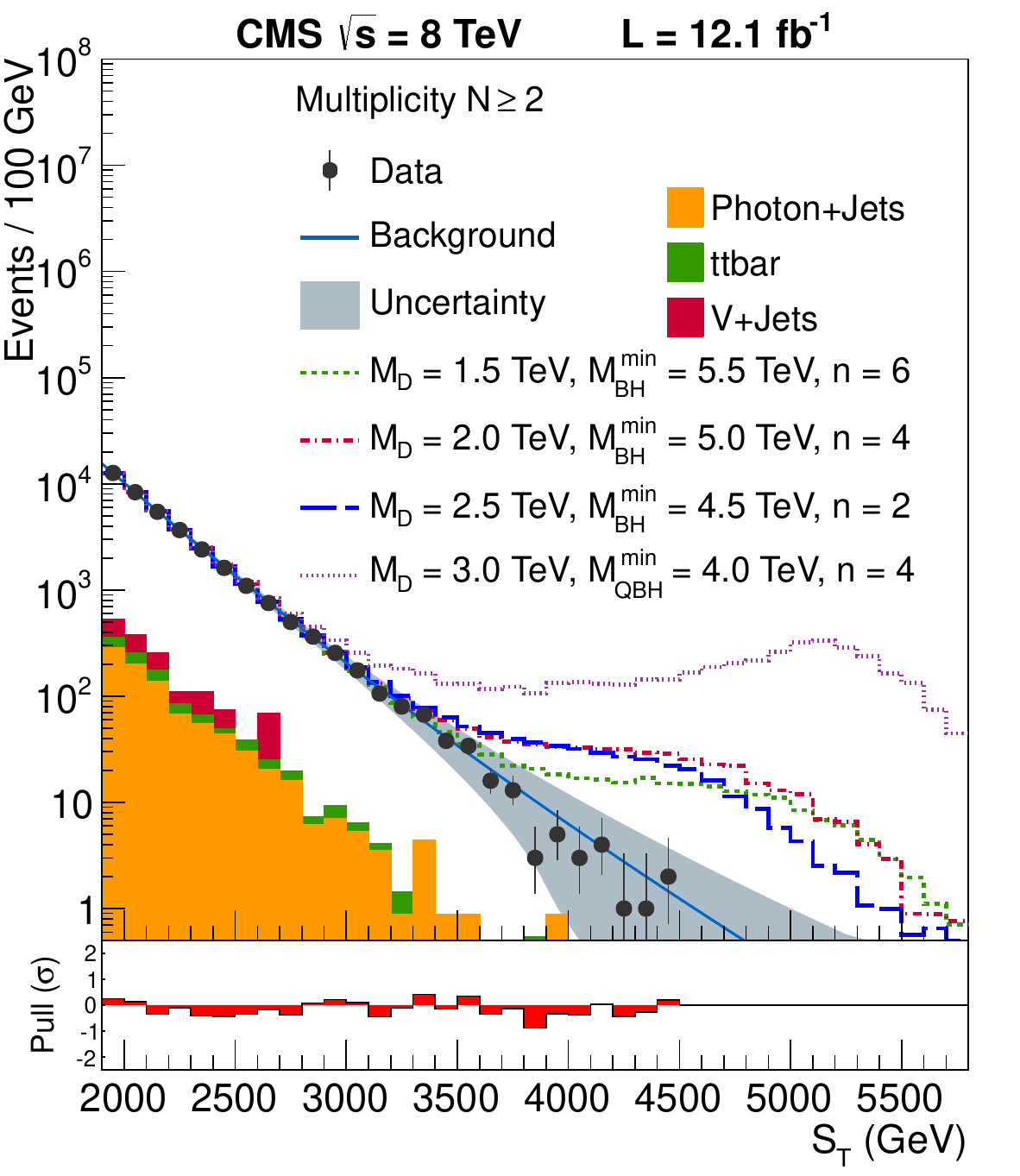}
\caption{Distribution of the scalar sum of transverse energy, $\ST$, for events with multiplicity: (Left) $N = 2$ and (right) $N \geq 2$ objects (photons, electrons, muons, or jets) in the final state. Observed data are depicted as points with statistical error bars; the solid line with a shaded band is the multijet background prediction from $N=2$ fit and its systematic uncertainty. Coloured histograms represent the $\gamma+\text{jets}$ (orange), $\PV+\text{jets}$ (red), and \ttbar (green)  backgrounds. Also shown are the expected semiclassical black hole signals for three parameter sets of the \textsc{BlackMax} nonrotating semiclassical black hole model, as well as a quantum black hole model. Here, $\MBHmin$ is the minimum black hole mass, $M_\mathrm{QBH}^{\min}$ is the minimum quantum black hole mass, $\MD$ is the multidimensional Planck scale, and $n$ is the number of extra dimensions. The bottom panels in each plot show the pull distribution (defined as ($\text{data} - \text{background})/\sigma(\text{data} - \text{background}$)) based on combined statistical and systematic uncertainty (dominated by the latter). Note that the systematic uncertainty is fully correlated bin-to-bin. Also shown in the $N = 2$ plot, is the background optimization based on a fit to $N=3$ data (dotted line). The difference between the $N=2$ and $N=3$ background fits are covered by the systematic uncertainty band used in the analysis.
}
\label{fig:STexclusive}
\end{figure}

The SM background is completely dominated by QCD multijet production and is estimated directly from data using a method based on $\ST$ multiplicity invariance~\cite{CMSBH,bh_full2011_jhep,Tsang:2011jta}. All other backgrounds are negligibly small in the $\ST$ range used in this analysis, as shown in figure~\ref{fig:STexclusive}. The multijet background estimation method is based on the empirical observation that the shape of the $\ST$ spectrum is approximately independent of $N$, so the shapes of the $\ST$ spectrum for any number of objects can be estimated using a fit to the dijet data ($N = 2$). The dijet mass spectrum has been previously studied in dedicated analyses~\cite{dijet2011,dijet2012}, as well as in the earlier searches for black holes at lower masses~\cite{CMSBH,bh_full2011_jhep} and is known not to exhibit any signal-like features in the range of $1.8 < \ST <2.8$\TeV, which is used to obtain the background shape. The central value of the background shape and its uncertainty are determined from the fit to several semi-empirical template functions~\cite{bh_full2011_jhep}, by taking the best fit function as the central value and the envelope of the alternative fits as the measure of systematic uncertainty in the background shape. The background shape is parameterized with the function $P_0(1 + x)^{P_1} /x^{P_2+P_3 \log(x)}$, and the uncertainty envelope is defined with two additional functions, $P_0/(P_1 + P_2 x + x^2)^{P_3}$ and $P_0/(P_1 + x)^{P_2}$. Here, $P_i$ are the fit parameters and $x = \ST/\sqrt{s} = \ST/8$\TeV. We also compare the fits to $N = 2$ data with fits to $N = 3$ data as a measure of the $\ST$ potential non-invariance of multiplicity. This effect is included in the total systematic uncertainty in the background prediction. Results of the fit can be seen in figure~\ref{fig:STexclusive}. 

The scaling of the background to higher multiplicities is performed by normalising the background shape to data in each inclusive multiplicity bin in the control range ($1.9 < \ST < 2.3$\TeV), where any significant signal contribution has been already ruled out by earlier analyses~\cite{CMSBH,bh_full2011_jhep}. The lower boundary of the control region is chosen to be substantially above the trigger and multiparticle ($N \times 50$\GeV) turn-on regions.

\begin{figure}[htbp]
\centering
\includegraphics[width=0.49\textwidth]{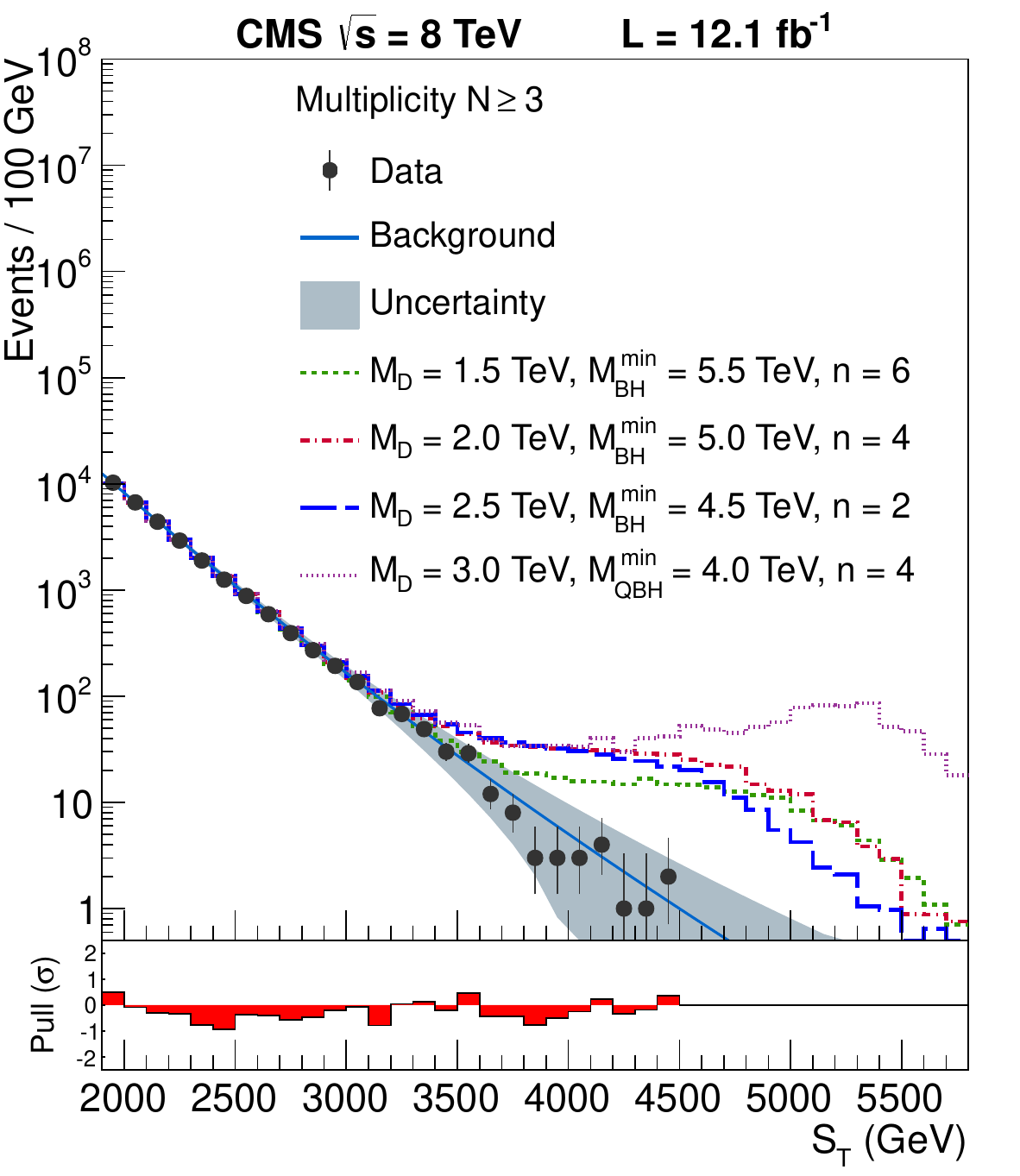}
\includegraphics[width=0.49\textwidth]{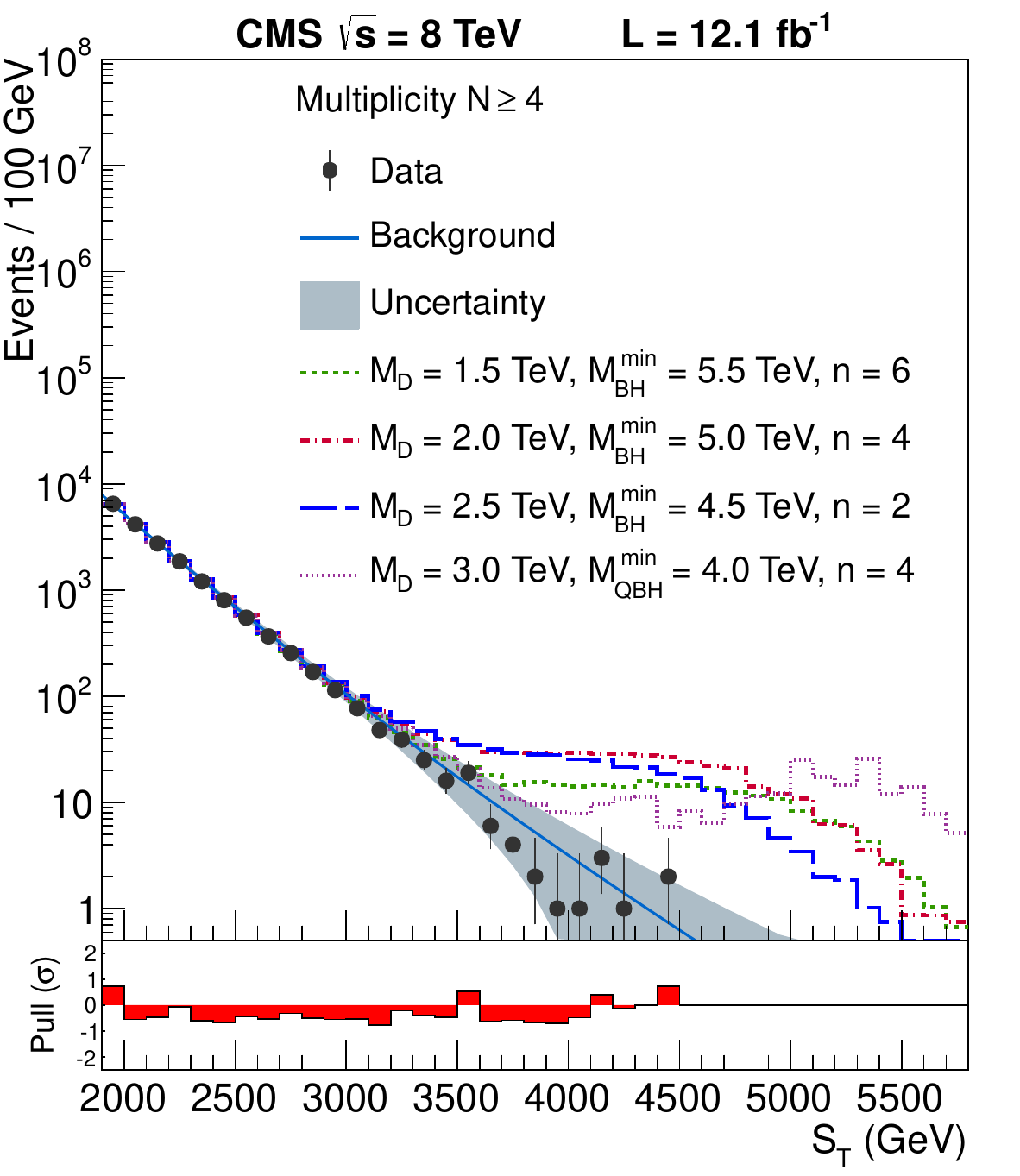}
\includegraphics[width=0.49\textwidth]{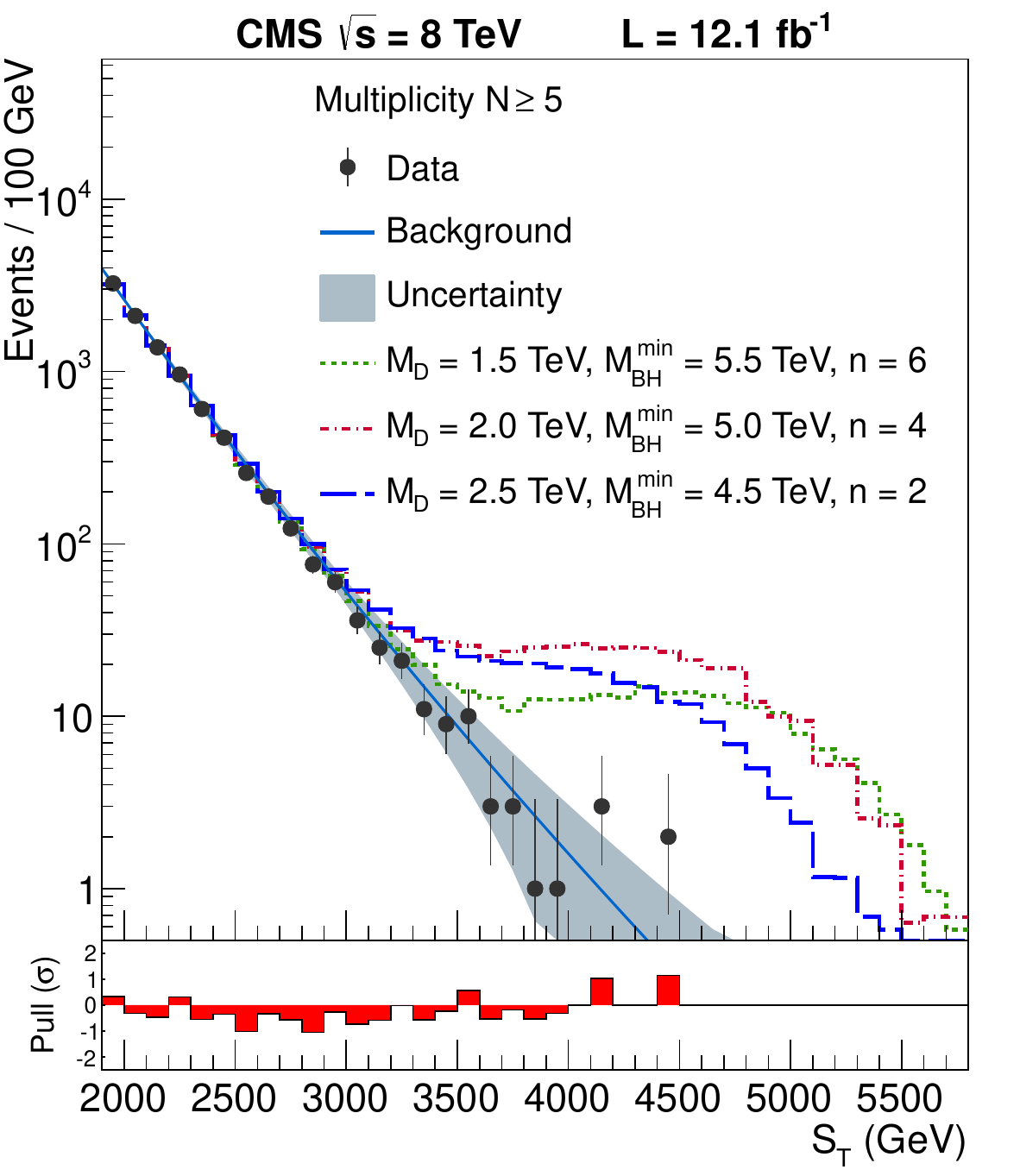}
\includegraphics[width=0.49\textwidth]{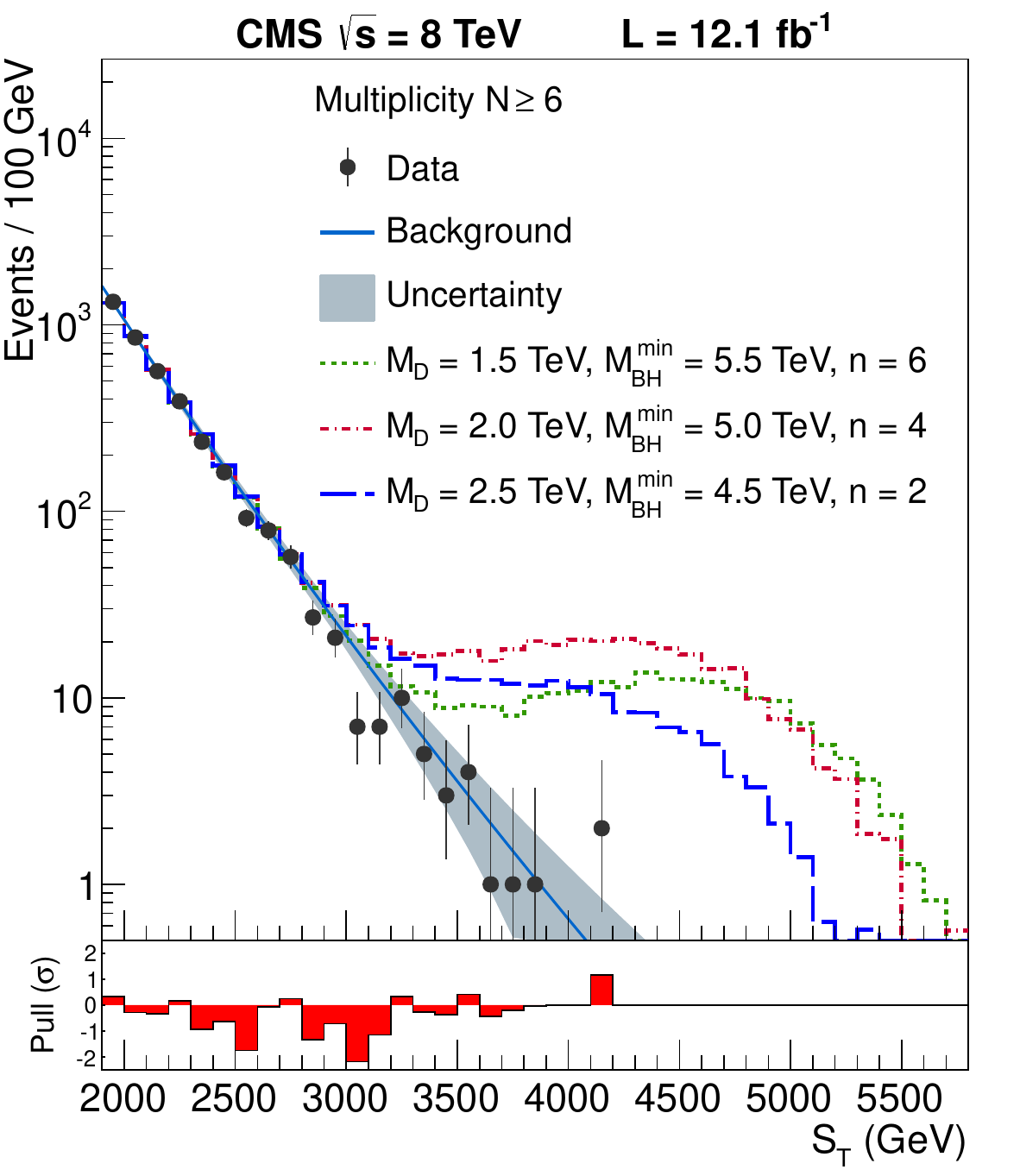}
\caption{Distribution of the scalar sum of transverse energy, $\ST$, for events with multiplicity: (Top left) $N \geq 3$, (top right) $N \geq 4$, (bottom left) $N \geq 5$, and (bottom right) $N \geq 6$ objects (photons, electrons, muons, or jets) in the final state. Observed data are depicted as points with statistical error bars; the solid line with a shaded band is the multijet background prediction and its systematic uncertainty. Also shown are the expected semiclassical black hole signals for three parameter sets of the \textsc{BlackMax} nonrotating black hole model, as well as a quantum black hole signal of the \textsc{qbh} model. Here, $\MBHmin$ is the minimum black hole mass, $M_\mathrm{QBH}^{\min}$ is the minimum quantum black hole mass, $\MD$ is the multidimensional Planck scale, and $n$ is the number of extra dimensions. The bottom panels in each plot show the pull distribution (defined as ($\text{data} - \text{background})/\sigma(\text{data} - \text{background}$)) based on combined statistical and systematic uncertainty (dominated by the latter). Note that the systematic uncertainty is fully correlated bin-to-bin.
}
\label{fig:STinclusive1}
\end{figure}

\begin{figure}[htbp]
\centering
\includegraphics[width=0.49\textwidth]{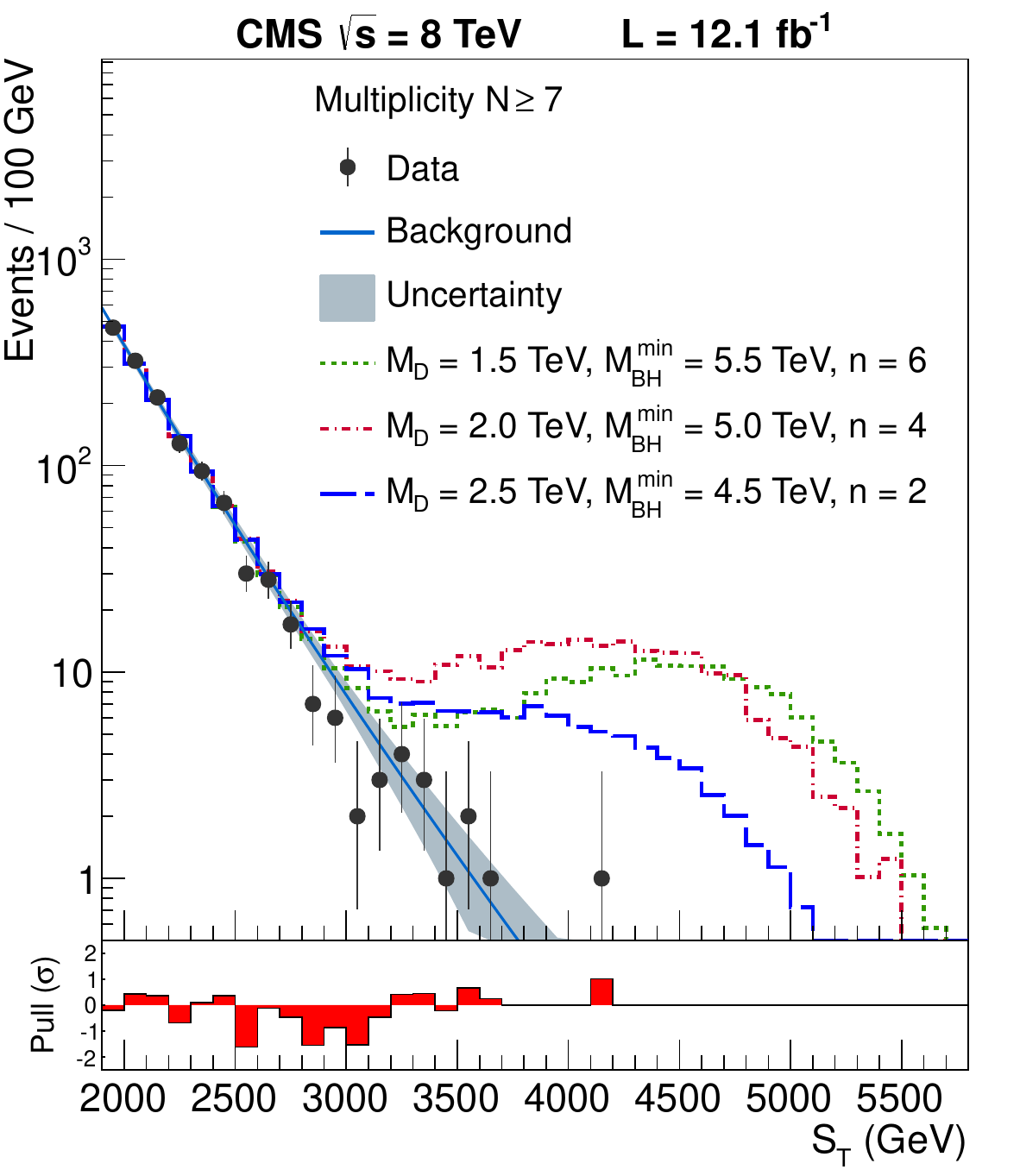}
\includegraphics[width=0.49\textwidth]{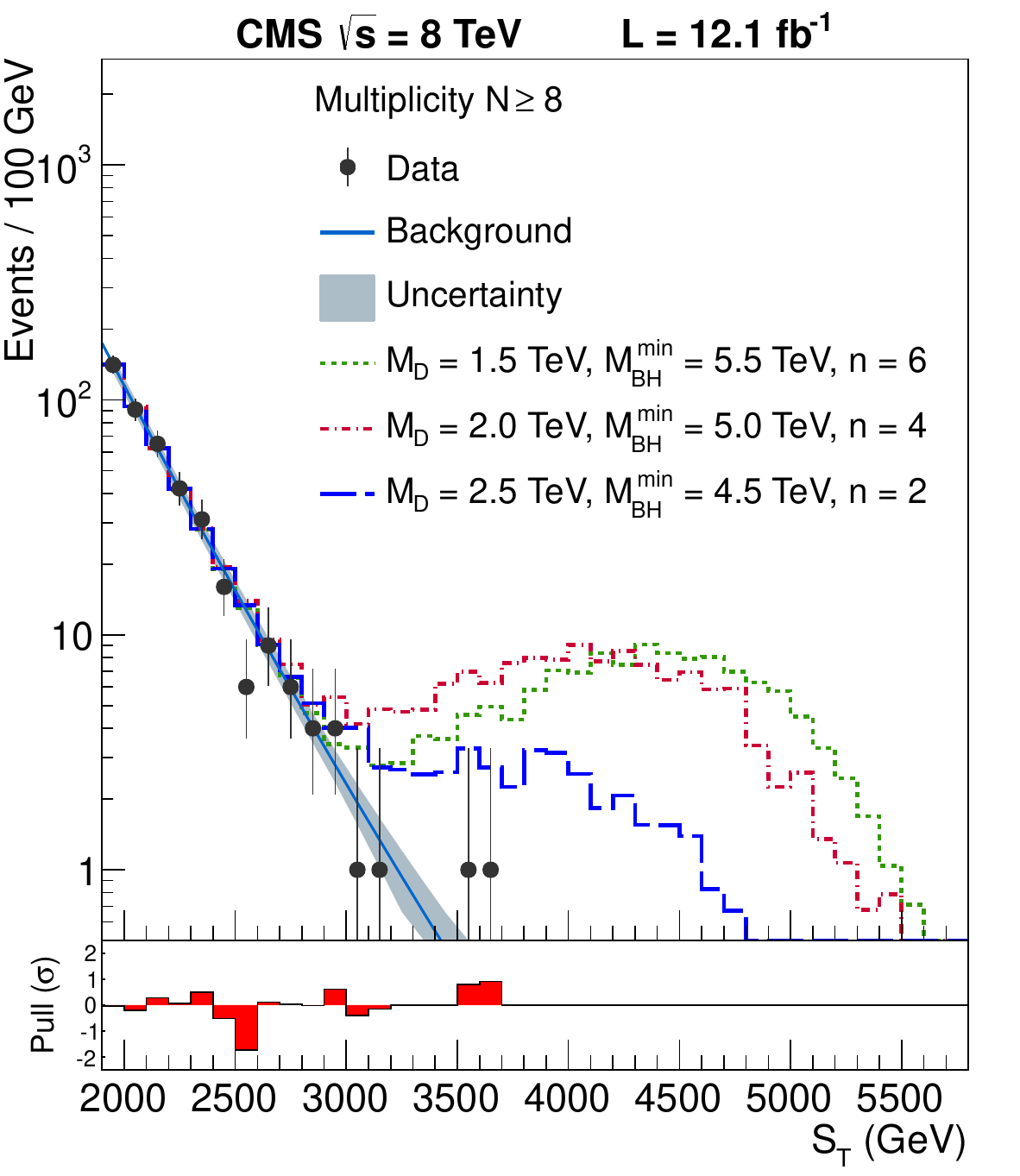}
\includegraphics[width=0.49\textwidth]{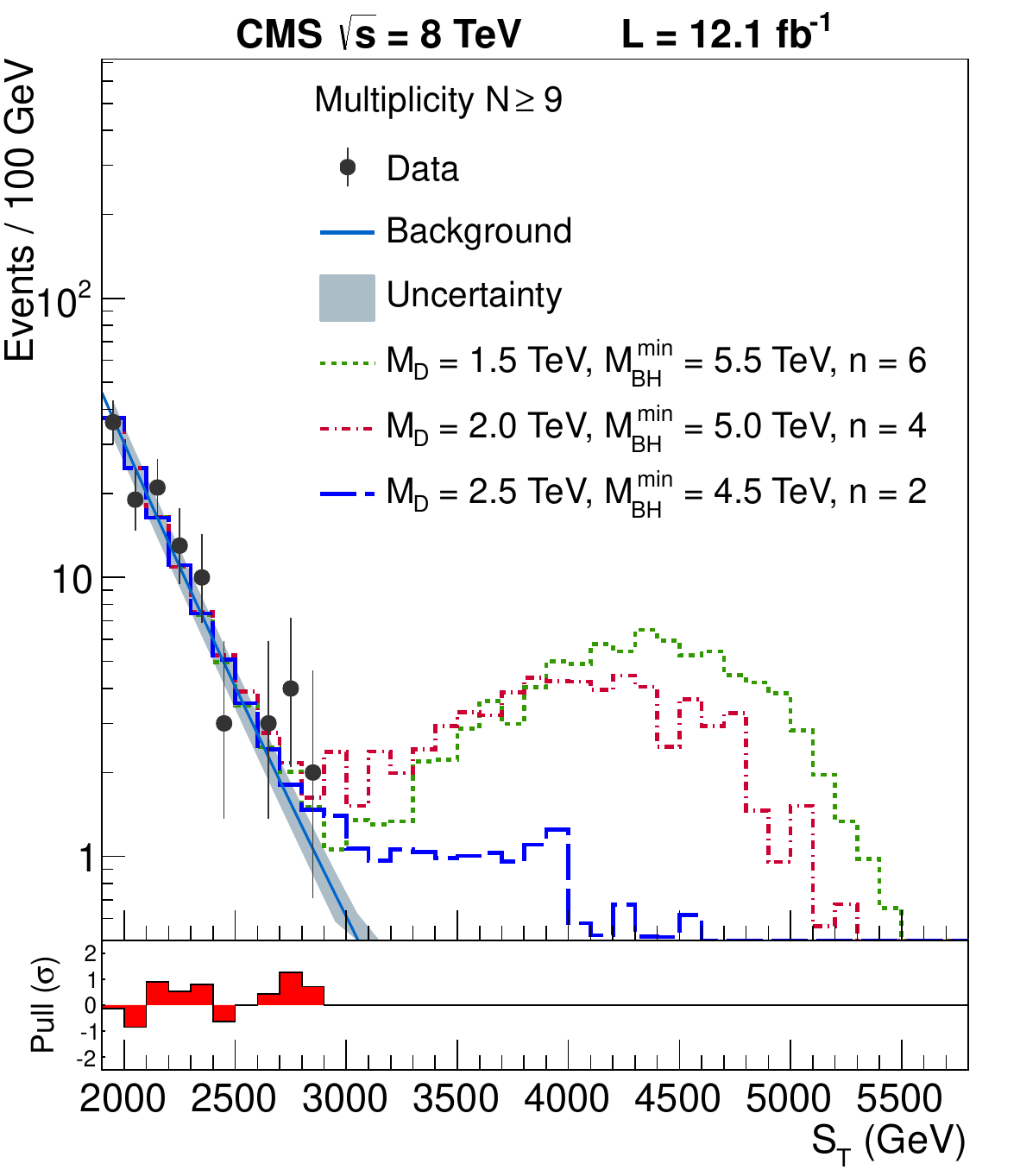}
\includegraphics[width=0.49\textwidth]{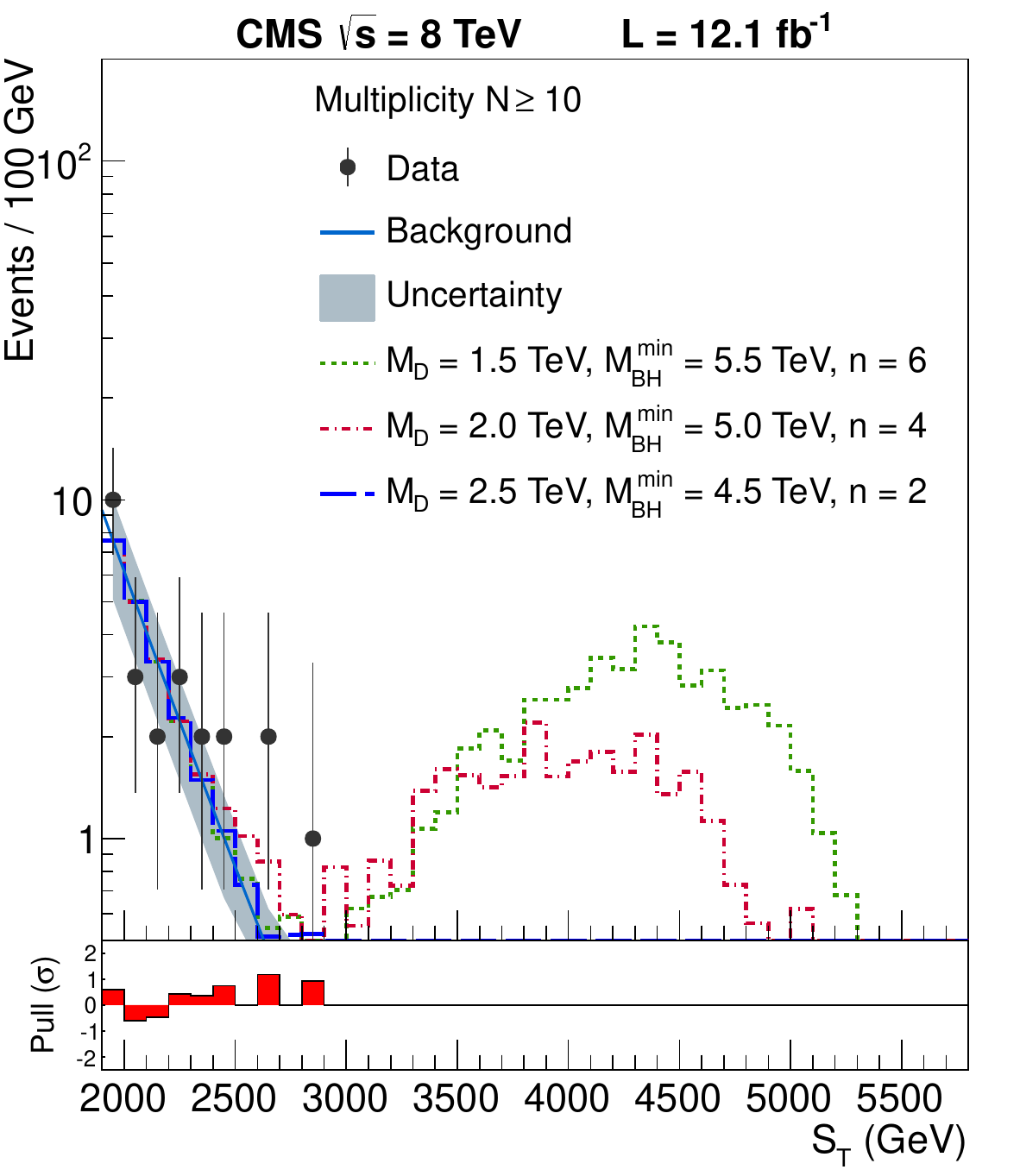}
\caption{Distribution of the scalar sum of transverse energy, $\ST$, for events with multiplicity: (Top left) $N \geq 7$, (top right) $N \geq 8$, (bottom left) $N \geq 9$, and (bottom right) $N \geq 10$ objects (photons, electrons, muons, or jets) in the final state. Observed data are depicted as points with statistical error bars; the solid line with a shaded band is the multijet background prediction and its systematic uncertainty. Also shown are the expected semiclassical black hole signals for three parameter sets of the \textsc{BlackMax} nonrotating black hole model. Here, $\MBHmin$ is the minimum black hole mass, $\MD$ is the multidimensional Planck scale, and $n$ is the number of extra dimensions. The bottom panels in each plot show the pull distribution (defined as ($\text{data} - \text{background})/\sigma(\text{data} - \text{background}$)) based on combined statistical and systematic uncertainty (dominated by the latter). Note that the systematic uncertainty is fully correlated bin-to-bin.
}
\label{fig:STinclusive2}
\end{figure}

The $\ST$ distributions for data, for predicted background, and for several semiclassical and quantum black hole signal benchmarks are shown in figures~\ref{fig:STexclusive}--\ref{fig:STinclusive2} for a number of exclusive and inclusive multiplicities. We do not plot the quantum black hole signal $\ST$ distributions for inclusive multiplicities of five or more, as the search for quantum black holes is not sensitive in higher inclusive multiplicity bins. No statistically significant excess of events over the expected background is observed in any of these spectra.

\section{Results\label{s:results}}

In the absence of an excess of data over the background prediction, we set limits on black hole and string ball production rates. The following systematic uncertainties are taken into account in the limit setting procedure. 

The total uncertainty in the background includes the uncertainty due to the choice of the fit function (including the uncertainties in the best-fit values of the parameters), the statistics in the normalization region, the uncertainty due to the choice of fit range, and the difference between the fits to $N = 2$ and $N = 3$ data as a measure of the potential non-invariance of $\ST$ with jet multiplicity. The normalization uncertainty is derived from the number of events in the normalization region in each jet multiplicity bin, and is negligible compared to the shape uncertainty, except for the $N \ge 10$ bin. The total uncertainty rises with $\ST$ from 5\% to as much as 200\% at very high values of $\ST$, where the background extrapolation is unreliable, owing to insufficient data in the control regions. Typical values of the background uncertainty are 5\% at $\ST = 2$ TeV, 18\% at $\ST = 3$ TeV, and 95\% at $\ST = 4$ TeV. The possible violation of $\ST$ invariance with jet multiplicity can be gauged from the bottom panes of figures~\ref{fig:STinclusive1} and \ref{fig:STinclusive2} and does not show any trends with increasing multiplicity. The effects of possible deviations from $\ST$ shape invariance are covered by the above systematic uncertainties in the fit. The uncertainties in the signal include the 8\% uncertainty due to the jet energy scale, which is known to $\approx 2\%$~\cite{JES}; a 6\% uncertainty in the signal acceptance due to the PDF choice, as determined using the prescribed PDF4LHC recipe~\cite{PDF4LHC}; and the 4.4\% uncertainty in the integrated luminosity~\cite{lumi}. As a result, the total systematic uncertainty in the signal is calculated to be 10\%. As the cross section for black hole production is known only approximately and is highly model-dependent, no theoretical uncertainty on the signal cross section is applied, as it is used merely as a benchmark.

\begin{figure}[htbp]
\centering
\includegraphics[width=0.45\textwidth]{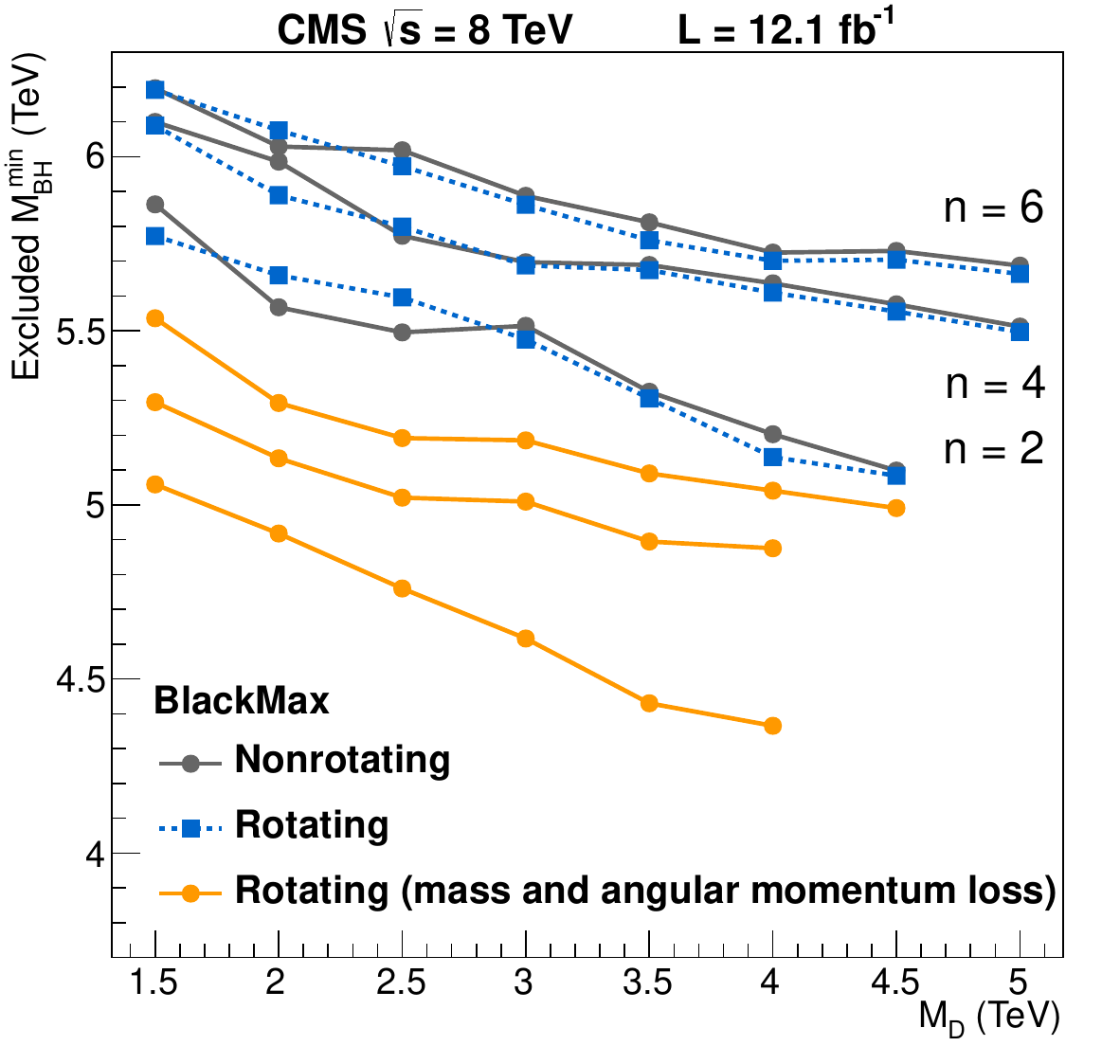}
\includegraphics[width=0.45\textwidth]{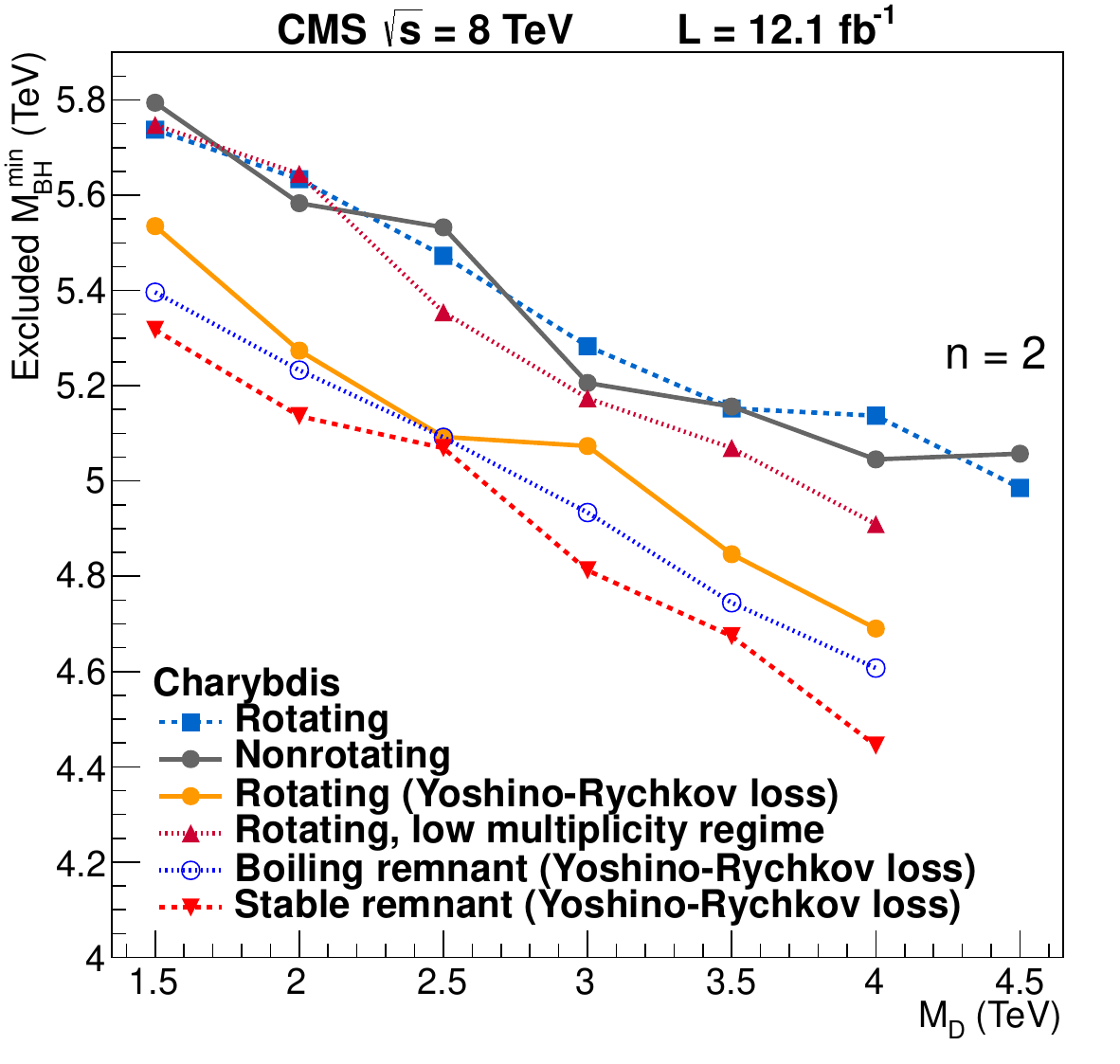}
\includegraphics[width=0.45\textwidth]{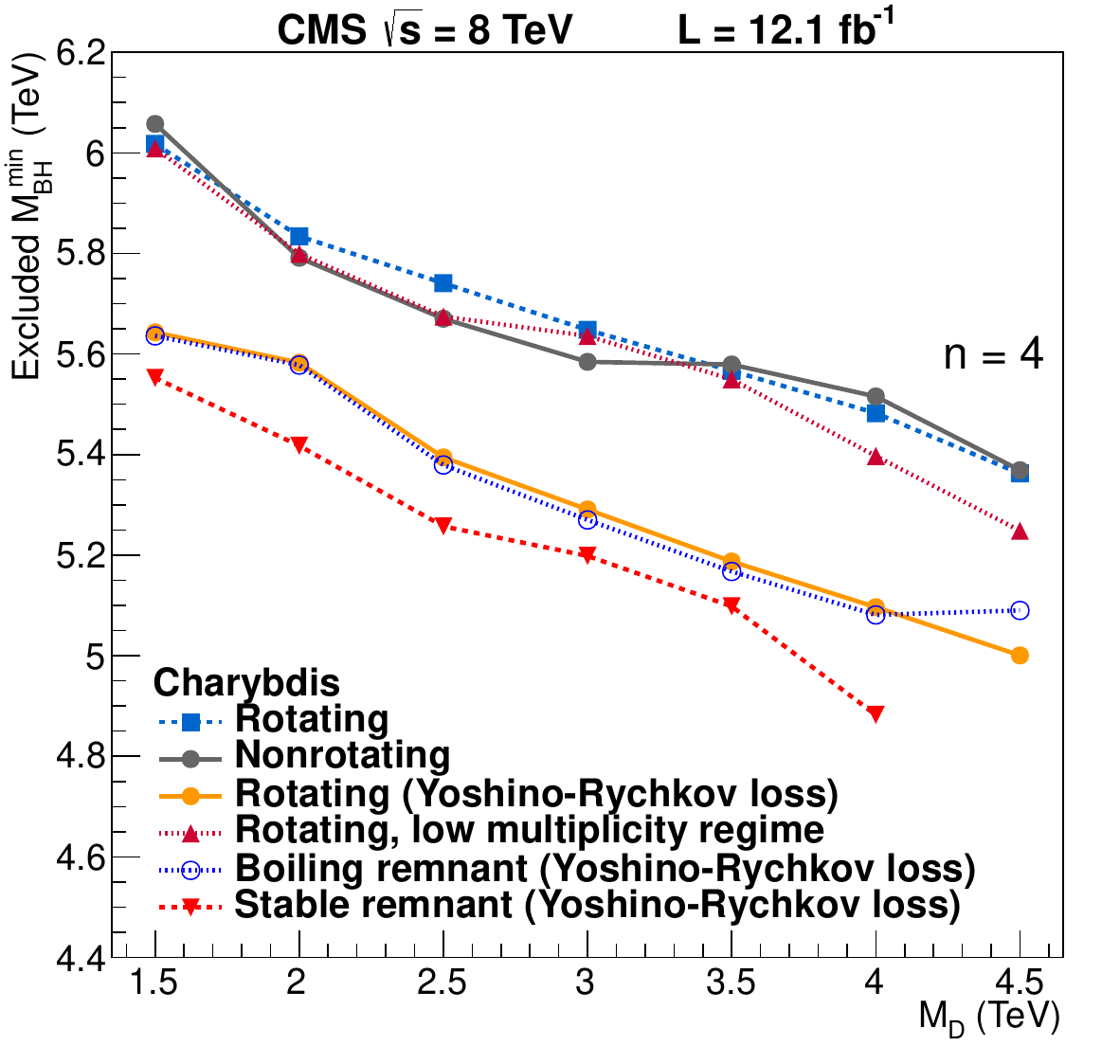}
\includegraphics[width=0.45\textwidth]{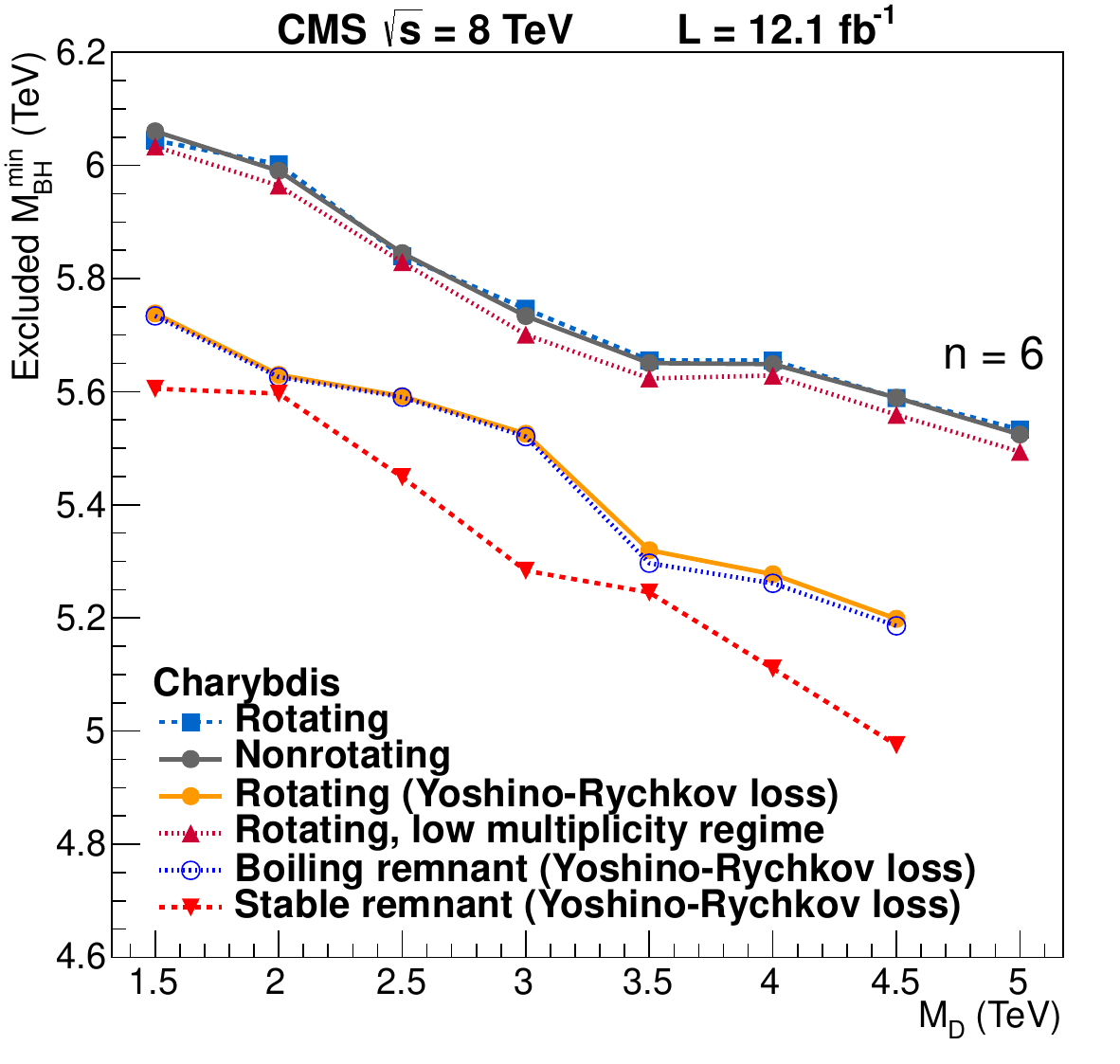}
\caption{The 95\% CL lower limits on the semiclassical black hole mass derived from the upper 95\% CL limits on cross section times branching fraction as a function of the fundamental Planck scale $\MD$, for various models.  The areas below each curve are excluded by this search. Top left: \textsc{BlackMax} black hole models without the stable remnant. Top right and bottom row: \textsc{Charybdis} black hole models with or without the stable remnant. The number $n$ of extra dimensions is labelled accordingly.}
\label{fig:BHpar}
\end{figure}

\begin{figure}[htbp]
\centering
\includegraphics[width=0.45\textwidth]{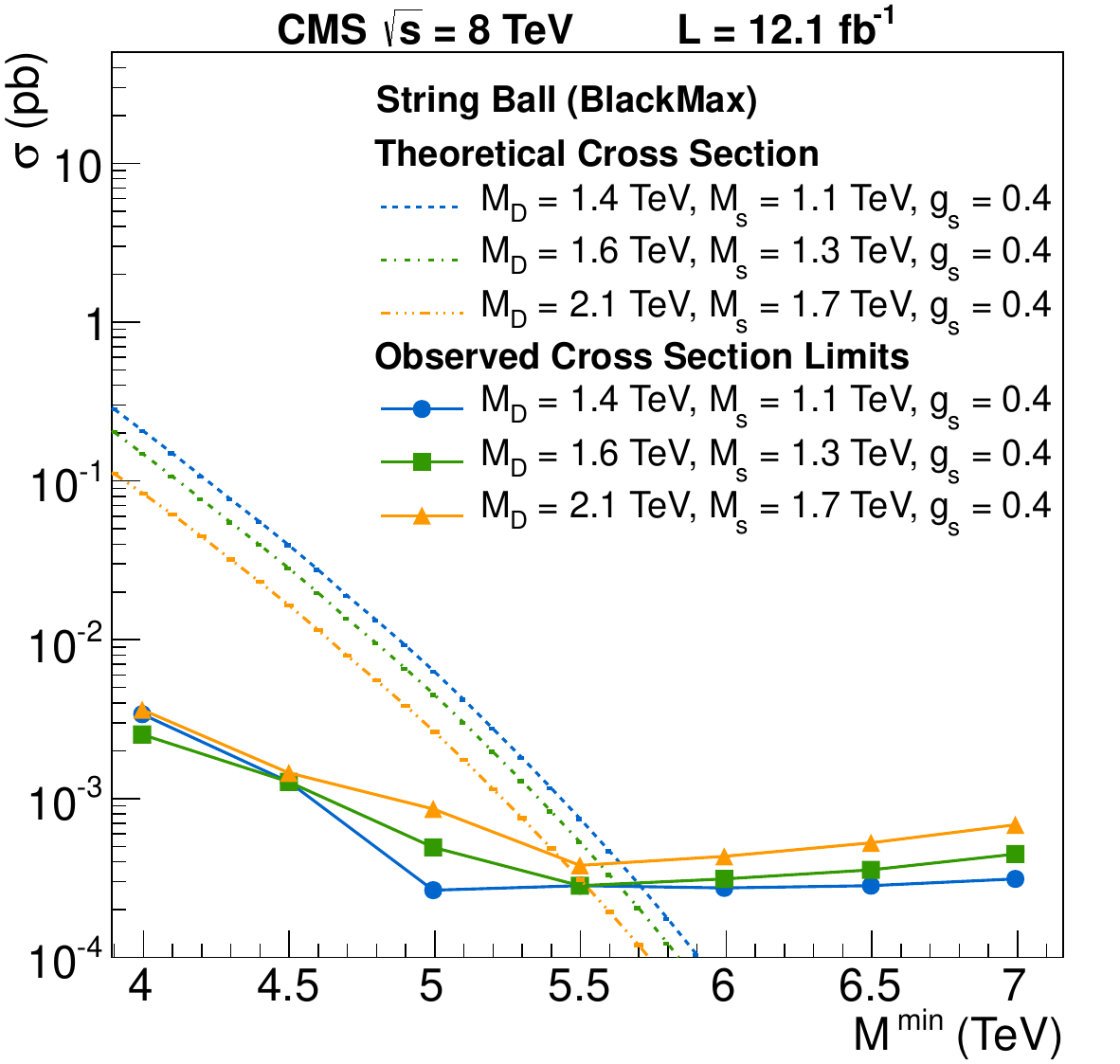}
\includegraphics[width=0.45\textwidth]{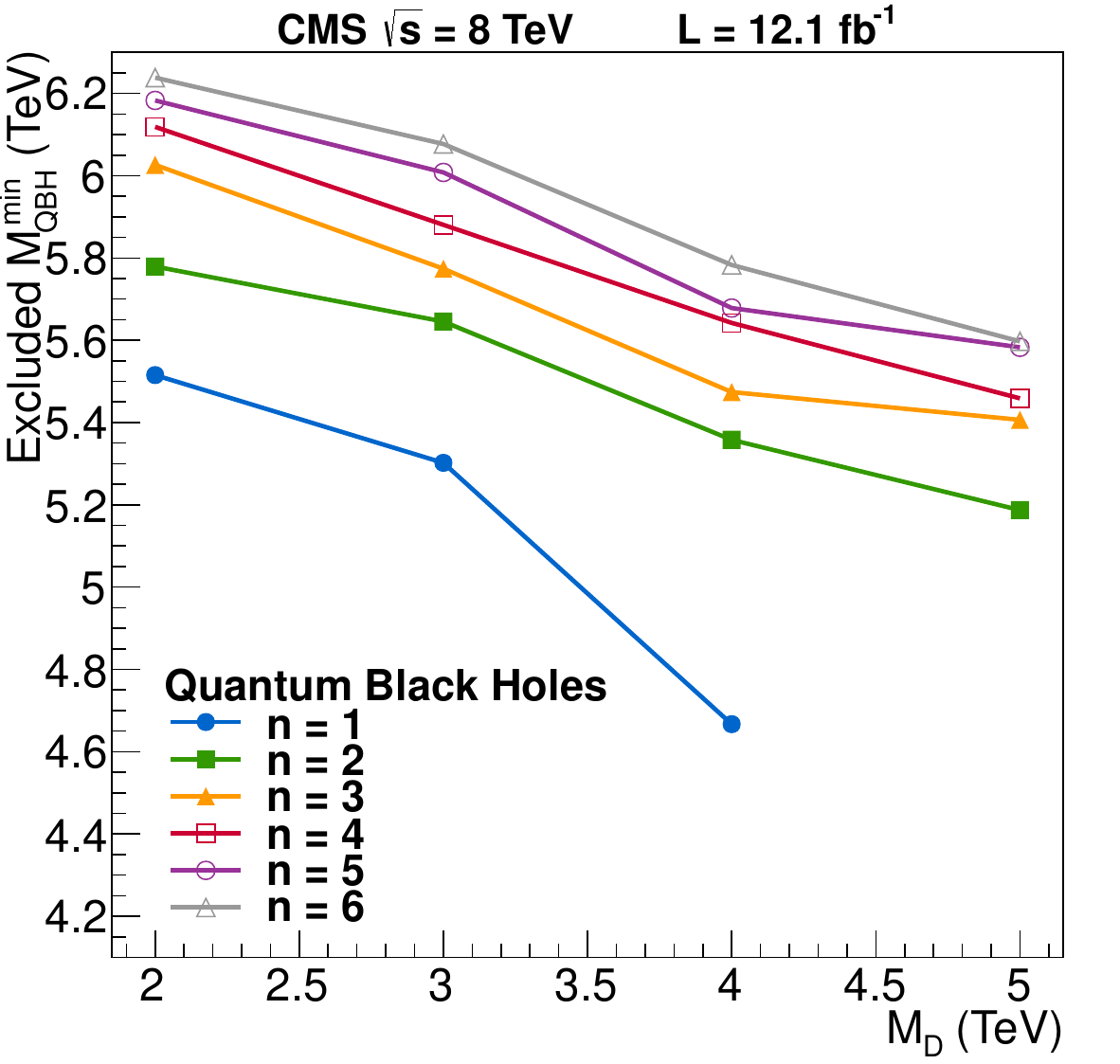}
\caption{(Left) The cross section upper limits at 95\% CL from the counting experiments optimized for various string ball parameter sets (solid lines) compared with predicted signal production cross section (dashed lines) as a function of minimum string ball mass. Here, $\MD$ is the multidimensional Planck scale, $M_\mathrm{S}$ is the string scale, and $g_\mathrm{S} = 0.4$ is the string coupling. (Right) Lower quantum black hole mass limits at $95\%$ CL as functions of the fundamental Planck scale $\MD$ for various \textsc{qbh} black hole models with a number $n$ of extra dimensions from one to six.}
\label{fig:SBpar}
\end{figure}

For each set of model parameters, a test statistic $S/\sqrt{S+B}$, where $S$ and $B$ are the numbers of signal and background events, is used to choose an optimal combination of minimum $\ST$ and multiplicity. Limits are then set using a modified frequentist \CLs method~\cite{cls,cls1} with a Poisson likelihood of the observed number of events, given the predicted background multiplied by the likelihoods of a set of measurements of the nuisance parameters that are related to various systematic uncertainties, modelled by log-normal distributions. Counting experiments are performed to set a 95\% confidence level (CL) cross section upper limit for each model used in this analysis. These limits can be interpreted in terms of lower mass limits on black holes (figure~\ref{fig:BHpar}) and string balls (figure~\ref{fig:SBpar}) that range from 4.3 to 6.2\TeV. The mass limit plots show lower mass limits for a number of benchmark models as a function of the fundamental Planck scale, $\MD$. The areas below each curve are excluded by this analysis. We note that the benchmarks used for semiclassical black holes are subject to large theoretical uncertainties and that the limits on the minimum black hole mass numerically close to $\MD$ can not be treated as theoretically reliable.

For quantum black holes, which are characterized by a low final-state multiplicity $N$, the limits come from the $N \ge 2$ samples. As the $N \ge 2$ sample largely overlaps with the sample used for the background shape determination ($N = 2$), we use the $N \ge 2$ sample only to set limits on quantum black holes with masses above the range used for the background fit, as can be seen in figure~\ref{fig:STexclusive}. Note that the $n = 1$ case for quantum black holes corresponds to the RS black holes~\cite{mr}. In this case, $\MD$ is the Planck scale times the exponential factor coming from the warping of the anti-deSitter space, and is expected to be of the order of the electroweak symmetry breaking scale, similar to the fundamental Planck scale in the ADD model. The limits on the quantum black holes mass are shown in figure~\ref{fig:SBpar}. All other benchmark model limits set in this paper correspond to the ADD model. The parameters used in simulations, the optimal combination of $\ST$ and multiplicity, signal acceptance, number of expected signal, observed, and background events, as well as observed and expected limits on the signal cross section are shown in Table~\ref{tab:BHT}.

\begin{table}[htbp]
\begin{center}
\caption{
Typical benchmark signal points for some of the models studied, corresponding leading-order cross sections $\sigma$, optimal selections on the minimum decay multiplicity ($N \geq N^{\min}$) and minimum $S_\mathrm{T}$, as well as signal acceptance $A$, expected number of signal events $N^\text{sig}$, number of observed events in data $N^\text{data}$, expected background $N^\text{bkg}$, and observed ($\sigma^{95}$) and expected ($\langle\sigma^{95}\rangle$) limits on the signal cross section at 95\% confidence level. Also here, $\MD$ is the multidimensinal Planck scale, $M_\mathrm{BH}$ is the minimum black hole mass, $M_\mathrm{QBH}$ is the minimum quantum black hole mass, $M_\mathrm{SB}$ is the minimum string ball mass, $M_\mathrm{S}$ is the string scale, $g_\mathrm{S}$ is the string coupling, and $n$ is the number of extra dimensions.}
\label{tab:BHT}
\begin{tabular}{|c|c|c|c|c|c|c|c|c|c|c|c|}
\hline
$\sigma$ & $N^{\min}$ & $S_\mathrm{T}^{\min}$ & $A$ & $N^\text{sig}$ & $N^\text{data}$ & $N^\text{bkg}$ & $\sigma^{95}$ & $\langle\sigma^{95}\rangle$ \\
(pb) & & (\TeVns{}) & & & & & (pb) & (pb)\\
\hline
\multicolumn{9}{|c|}{\textsc{BlackMax} nonrotating BH with $\MD = 2.5$\TeV, $M_\mathrm{BH} = 4.5$\TeV, and $n = 4$} \\ \hline
0.15 & 3 & 3.2 & 0.74 & 1338  &  213 & $228 \pm 111$ & $1.3 \times 10^{-2}$ & $(1.3 \pm 0.5)\times 10^{-2}$ \\ \hline
\multicolumn{9}{|c|}{\textsc{Charybdis} nonrotating BH w/ boiling remnant; $\MD = 1.5$\TeV, $M_\mathrm{BH} = 4.5$\TeV, and $n = 6$} \\ \hline
0.23 & 4 & 3.0 &   0.76 & 2056 &   244 & $290 \pm 99$ & $1.0 \times 10^{-2}$ & $(1.3 \pm 0.4)\times 10^{-2}$ \\ \hline
\multicolumn{9}{|c|}{\textsc{BlackMax} rotating BH; $\MD = 2.0$\TeV, $M_\mathrm{BH} = 5.5$\TeV, and $n = 6$} \\ \hline
0.01 & 3 & 4.0 &  0.59 & 71.2 &   11 & $15.6^{+22.6}_{-15.6}$ & $1.9\times 10^{-3}$ & $(2.0 \pm 0.6) \times 10^{-3}$\\ \hline
\multicolumn{9}{|c|}{\textsc{BlackMax} rotating BH w/ mass loss; $\MD = 3.0$\TeV, $M_\mathrm{BH} = 5.0$\TeV, and $n = 4$} \\ \hline
$1.4\times 10^{-3}$ & 3 & 4.2 & 0.41 & 7.1 &  4 & $8.2^{+15.1}_{-8.2}$ & $1.4\times 10^{-3}$ & $(1.5 \pm 0.6)\times 10^{-3}$\\ \hline
\multicolumn{9}{|c|}{\textsc{BlackMax} SB; $\MD = 2.1$\TeV, $M_\mathrm{SB} = 4.0$\TeV, $M_\mathrm{S} = 1.7$\TeV, and $g_\mathrm{S} = 0.4$} \\ \hline
0.08 & 6 & 2.8 & 0.65 & 656 &  89 & $123 \pm 29$  & $3.6\times 10^{-3}$ & $(5.0 \pm 1.9)\times 10^{-3}$\\ \hline
\multicolumn{9}{|c|}{\textsc{qbh} quantum BH; $\MD = 2.0$\TeV, $M_\mathrm{QBH} = 4.0$\TeV, and $n = 4$} \\ \hline
1.50 & 2 & 2.8 & 0.67 & 1211 &  1168 & $1180 \pm 274$ & $5.0\times 10^{-2}$ &  $(5.0\pm 1.7)\times 10^{-2}$ \\
\hline
\end{tabular}
\end{center}
\end{table}

\begin{figure}[htbp]
\centering
   \includegraphics[height=0.45\textwidth]{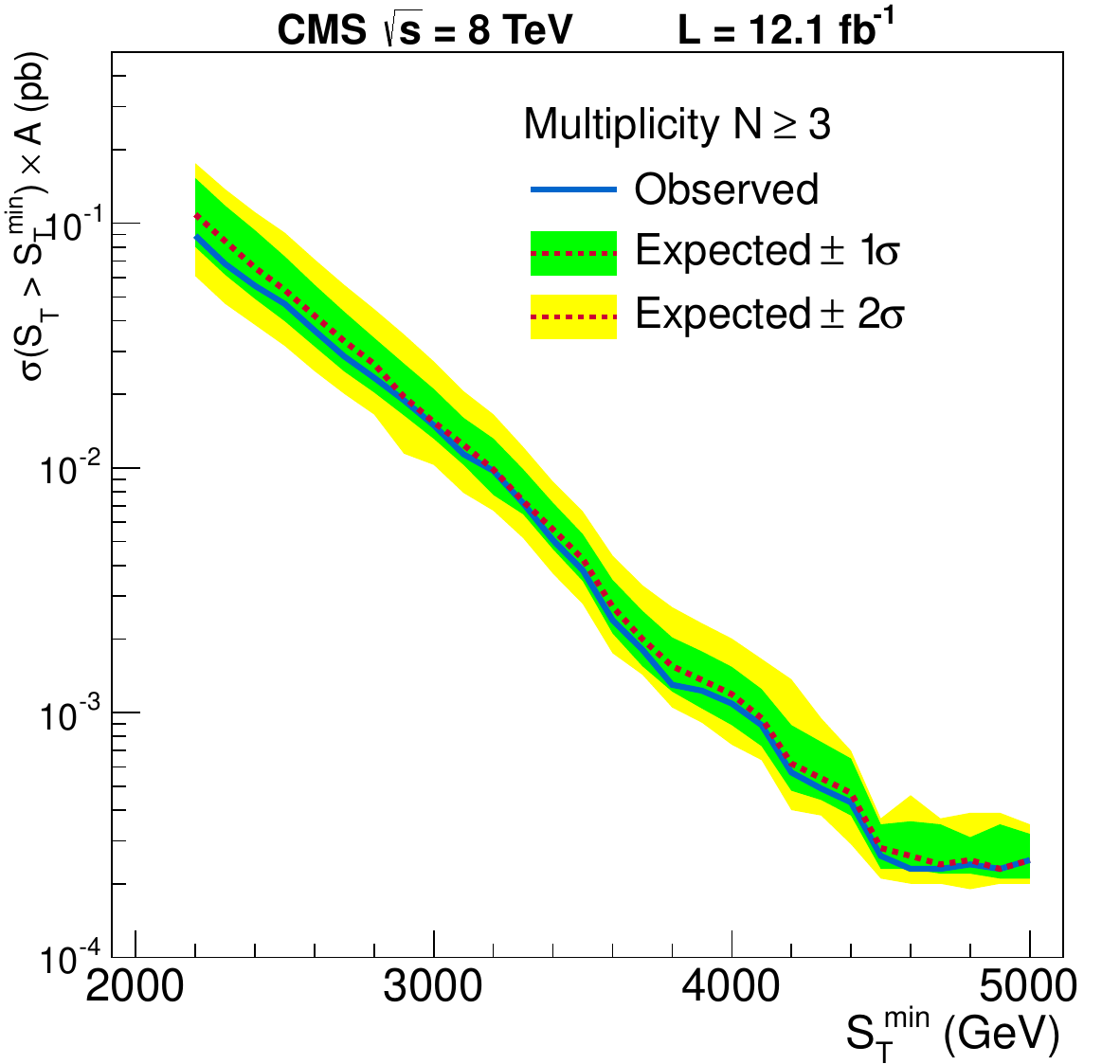}
   \includegraphics[height=0.45\textwidth]{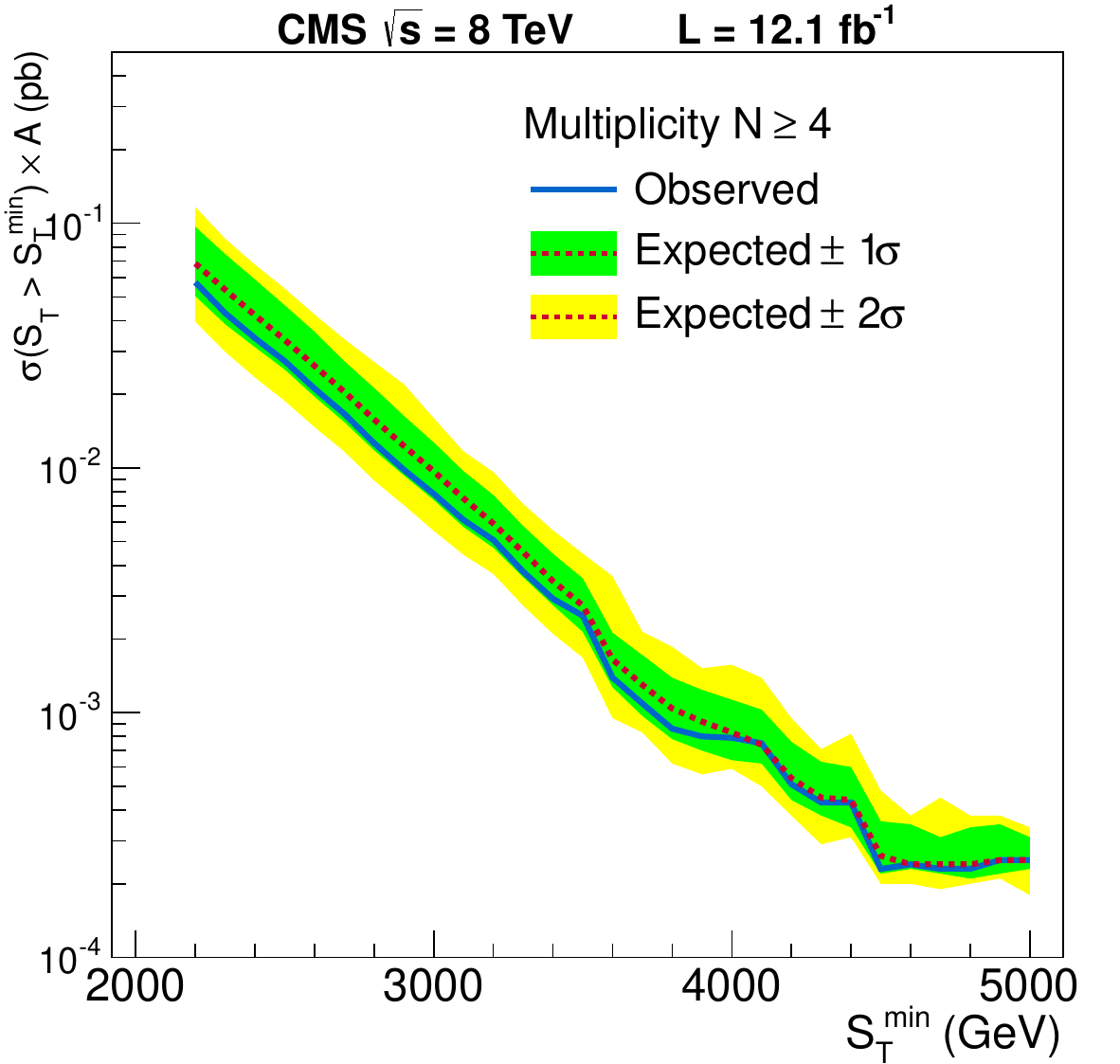}
   \includegraphics[height=0.45\textwidth]{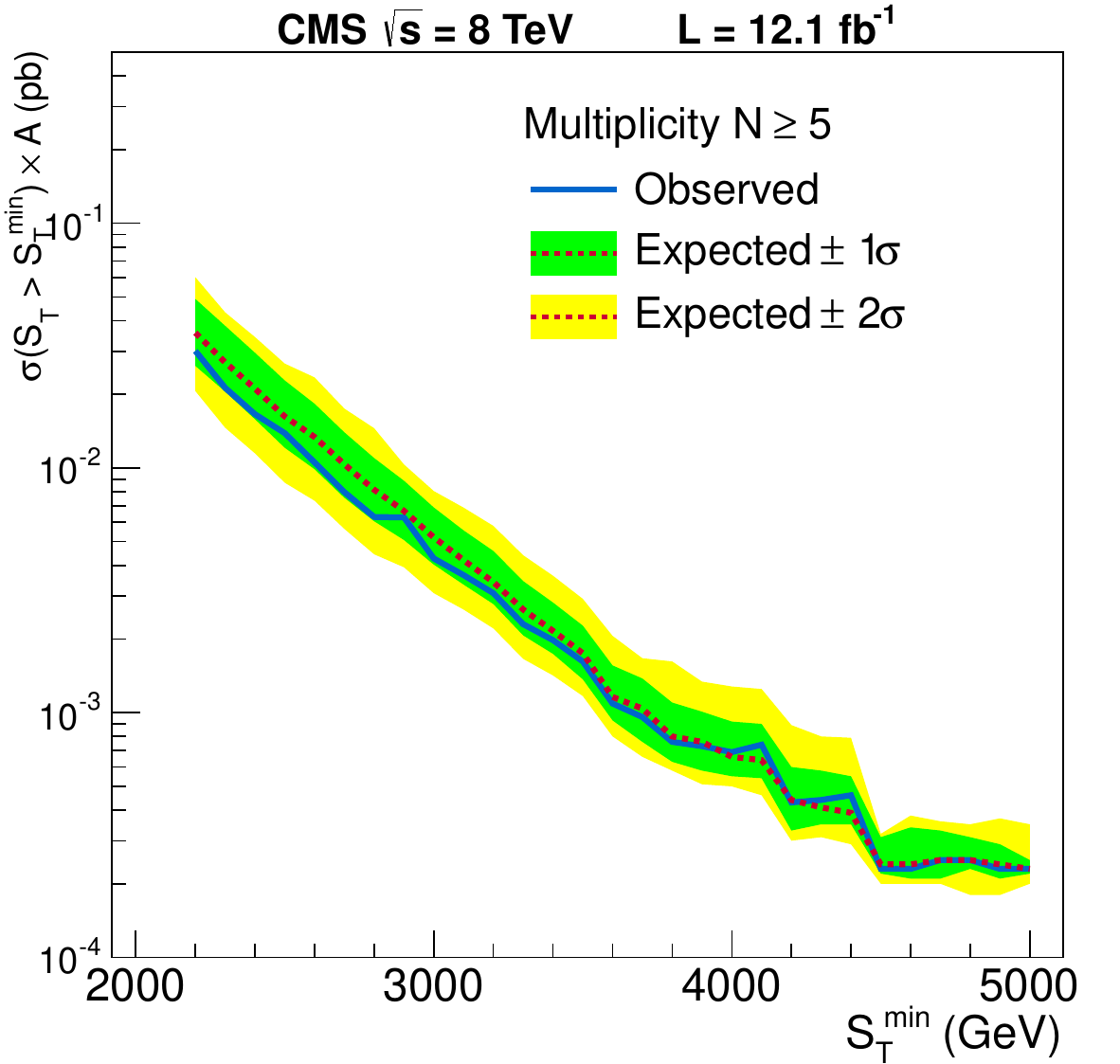}
   \includegraphics[height=0.45\textwidth]{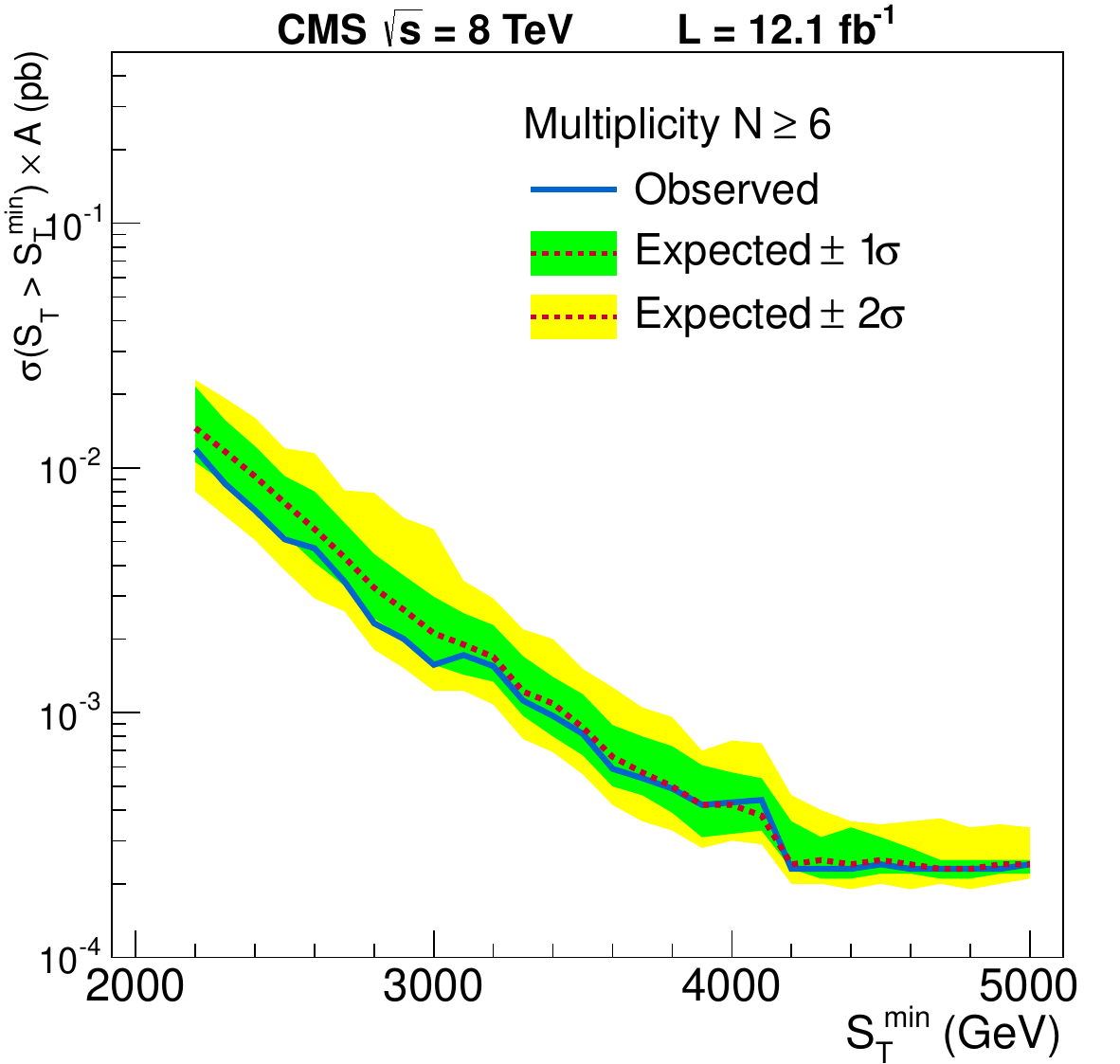}
\caption{Model-independent $95\%$ CL cross section times acceptance ($A$) upper limits for counting experiments with $\ST > \ST^{\min}$ as a function of $\ST^{\min}$ for events with multiplicity: (Top left) $N \ge 3$, (top right) $N \ge 4$, (bottom left) $N \ge 5$, and (bottom right) $N \ge 6$. The blue solid (red dotted) lines correspond to an observed (expected) limit for nominal signal acceptance uncertainty of $10\%$. The green (dark) and yellow (light) bands represent one and two standard deviations from the expected limits.}
\label{fig:modind_limits1}
\end{figure}

\begin{figure}[htbp]
\centering
   \includegraphics[height=0.45\textwidth]{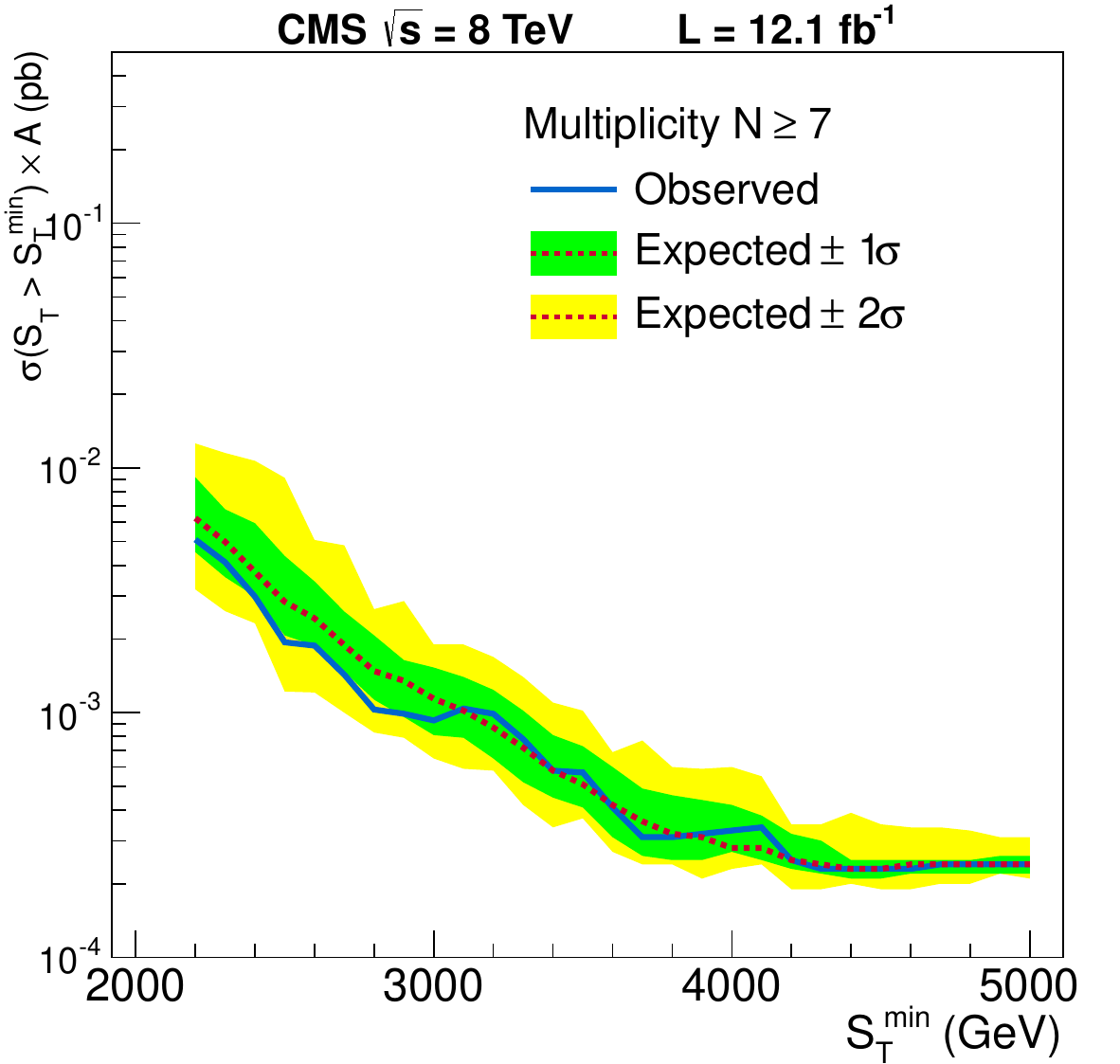}
   \includegraphics[height=0.45\textwidth]{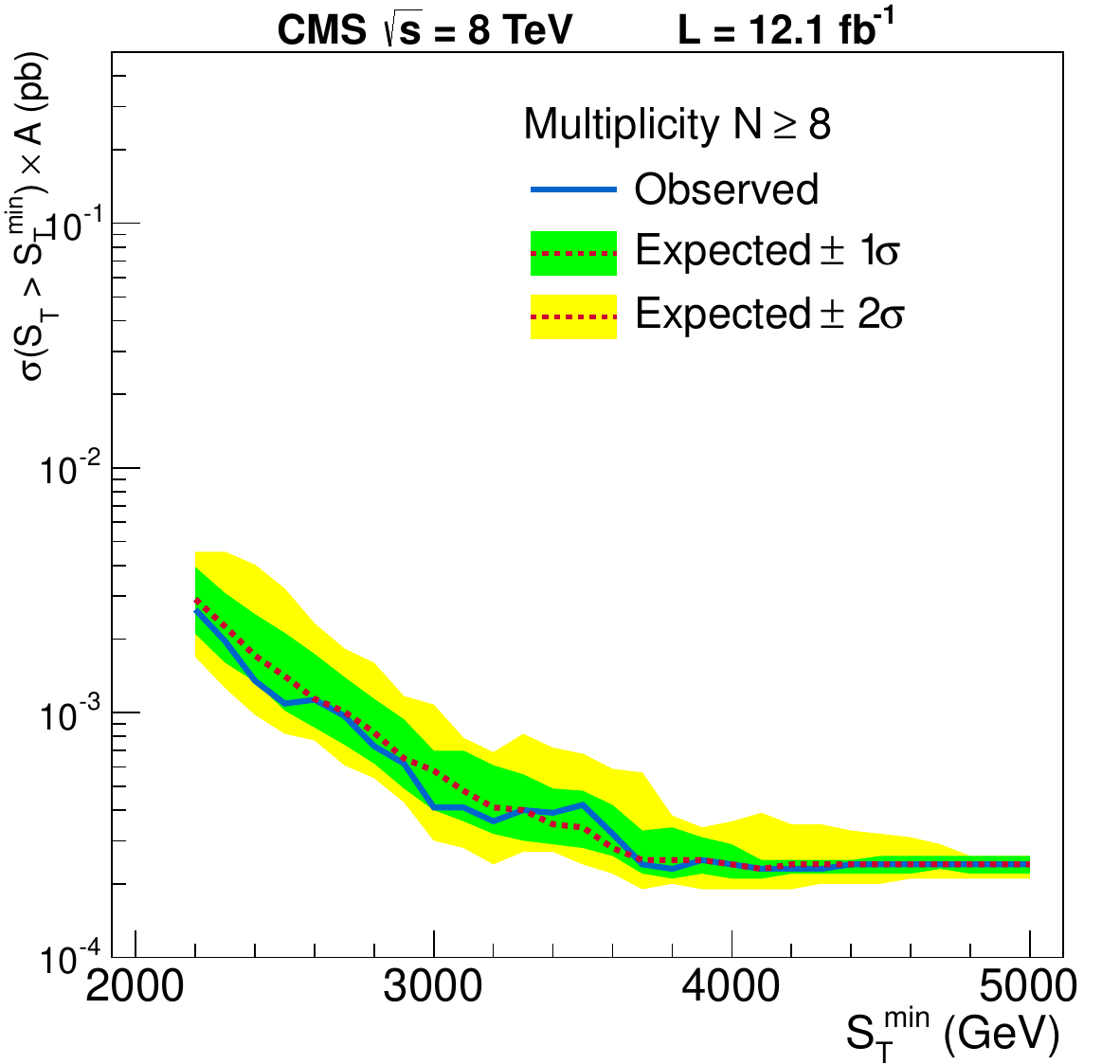}
   \includegraphics[height=0.45\textwidth]{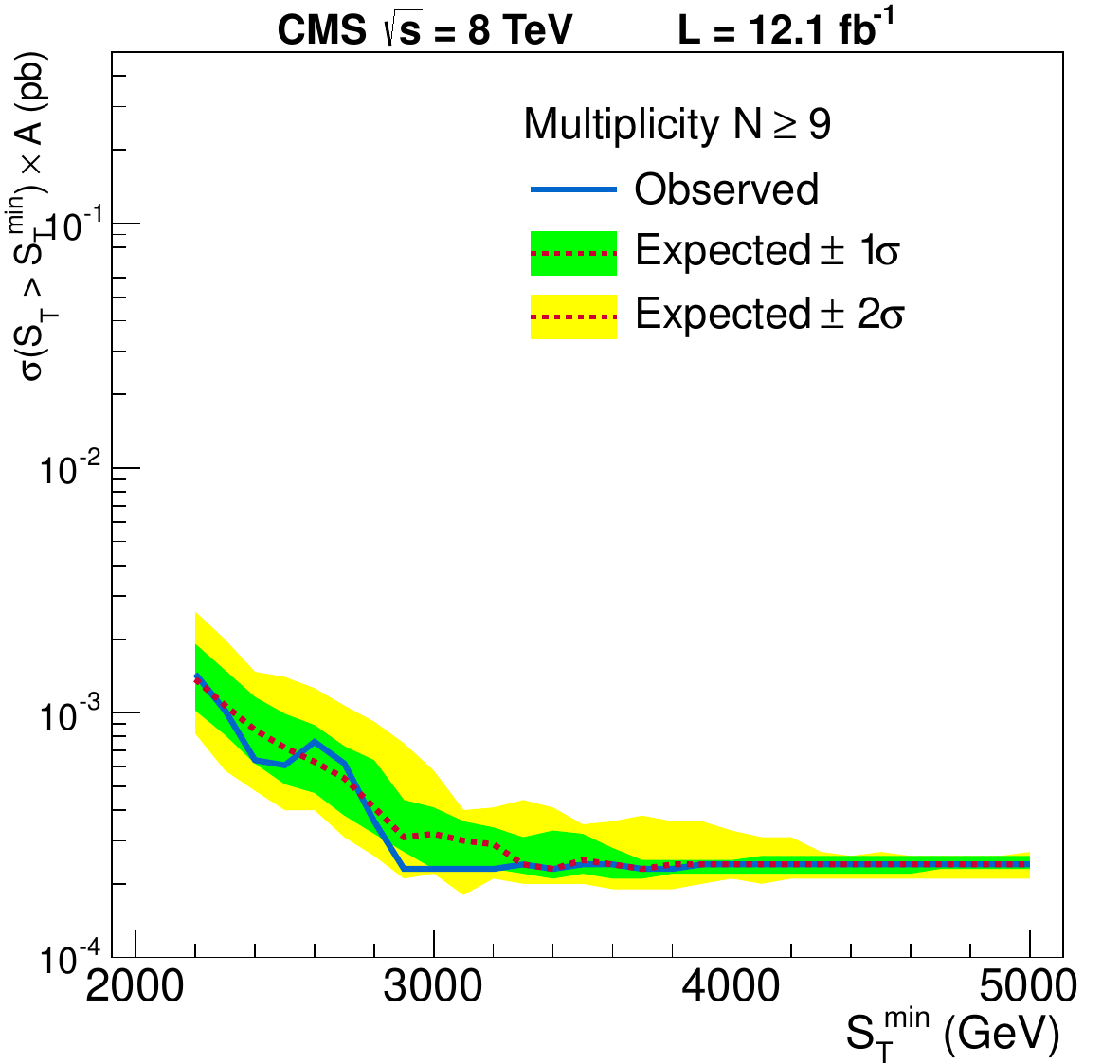}
   \includegraphics[height=0.45\textwidth]{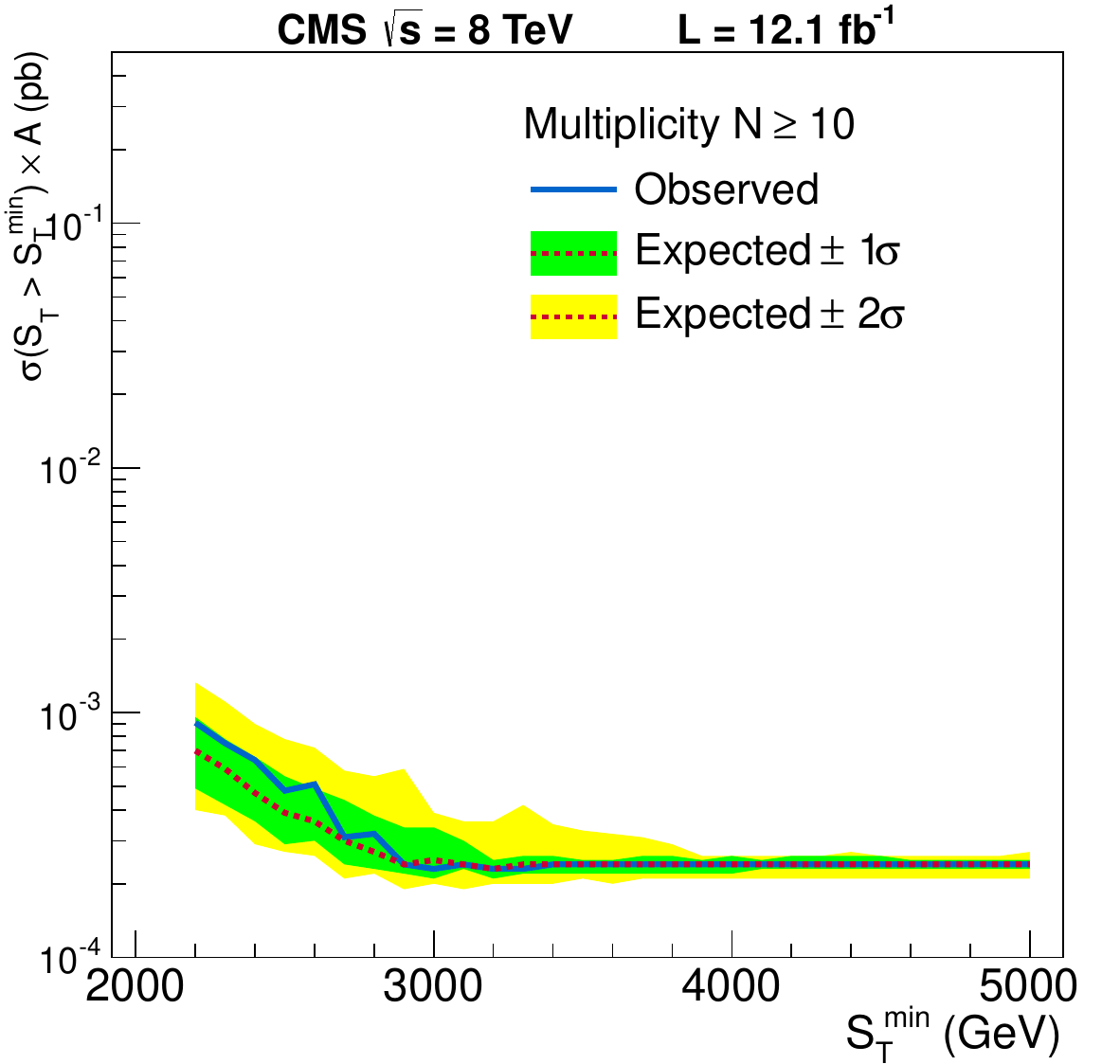}
\caption{Model-independent $95\%$ CL cross section times acceptance ($A$) upper limits for counting experiments with $\ST > \ST^{\min}$ as a function of $\ST^{\min}$ for events with multiplicity: (Top left) $N \ge 7$, (top right) $N \ge 8$, (bottom left) $N \ge 9$, and (bottom right) $N \ge 10$. The blue solid (red dotted) lines correspond to an observed (expected) limit for nominal signal acceptance uncertainty of 10\%. The green (dark) and yellow (light) bands represent one and two standard deviations from the expected limits.}
\label{fig:modind_limits2}
\end{figure}

To extend the scope of this search, model-independent limits on the cross section times the acceptance ($A$) are computed for high-$\ST$ inclusive final states for $N \geq 3, 4, 5, 6, 7$, $8$, $9$, and $10$, as a function of minimum $\ST$ (figures~\ref{fig:modind_limits1},~\ref{fig:modind_limits2}). The intersection of these limits with theoretical predictions for the cross section within the fiducial and kinematic selections used in this analysis could be used to constrain other models of new physics resulting in energetic, multiparticle final states. These model-independent limits on the cross section times acceptance are as low as 0.2\unit{fb} at 95\% CL for minimum $\ST$ values above ${\sim}4.5$\TeV, where no data events are observed.

\section{Summary\label{s:summary}}
A search for microscopic black holes and string balls at the LHC has been conducted, using a data sample corresponding to an integrated luminosity of $12.1 \pm 0.5$\fbinv of $\sqrt{s} = 8$\TeV pp collisions collected with the CMS detector at the LHC in 2012. Comparing the distributions of the scalar sum of the transverse momenta of all the final-state objects in data events with those from the estimated background, new model-independent limits are set that can be used to constrain a wide variety of models. With this search, semiclassical and quantum black holes with masses below 4.3--6.2\TeV are excluded in the context of a number of benchmark models. Stringent limits on black hole precursors\,--\,string balls\,--\,are also set. These limits extend significantly the previously probed regime of black hole production at hadron colliders and represent the most restrictive exclusions on these objects to date.

\section*{Acknowledgements}

We congratulate our colleagues in the CERN accelerator departments for the excellent performance of the LHC and thank the technical and administrative staffs at CERN and at other CMS institutes for their contributions to the success of the CMS effort. In addition, we gratefully acknowledge the computing centres and personnel of the Worldwide LHC Computing Grid for delivering so effectively the computing infrastructure essential to our analyses. Finally, we acknowledge the enduring support for the construction and operation of the LHC and the CMS detector provided by the following funding agencies: BMWF and FWF (Austria); FNRS and FWO (Belgium); CNPq, CAPES, FAPERJ, and FAPESP (Brazil); MEYS (Bulgaria); CERN; CAS, MoST, and NSFC (China); COLCIENCIAS (Colombia); MSES (Croatia); RPF (Cyprus); MoER, SF0690030s09 and ERDF (Estonia); Academy of Finland, MEC, and HIP (Finland); CEA and CNRS/IN2P3 (France); BMBF, DFG, and HGF (Germany); GSRT (Greece); OTKA and NKTH (Hungary); DAE and DST (India); IPM (Iran); SFI (Ireland); INFN (Italy); NRF and WCU (Republic of Korea); LAS (Lithuania); CINVESTAV, CONACYT, SEP, and UASLP-FAI (Mexico); MSI (New Zealand); PAEC (Pakistan); MSHE and NSC (Poland); FCT (Portugal); JINR (Armenia, Belarus, Georgia, Ukraine, Uzbekistan); MON, RosAtom, RAS and RFBR (Russia); MSTD (Serbia); SEIDI and CPAN (Spain); Swiss Funding Agencies (Switzerland); NSC (Taipei); ThEPCenter, IPST and NSTDA (Thailand); TUBITAK and TAEK (Turkey); NASU (Ukraine); STFC (United Kingdom); DOE and NSF (USA).

\bibliography{auto_generated}   
\cleardoublepage \appendix\section{The CMS Collaboration \label{app:collab}}\begin{sloppypar}\hyphenpenalty=5000\widowpenalty=500\clubpenalty=5000\textbf{Yerevan Physics Institute,  Yerevan,  Armenia}\\*[0pt]
S.~Chatrchyan, V.~Khachatryan, A.M.~Sirunyan, A.~Tumasyan
\vskip\cmsinstskip
\textbf{Institut f\"{u}r Hochenergiephysik der OeAW,  Wien,  Austria}\\*[0pt]
W.~Adam, T.~Bergauer, M.~Dragicevic, J.~Er\"{o}, C.~Fabjan\cmsAuthorMark{1}, M.~Friedl, R.~Fr\"{u}hwirth\cmsAuthorMark{1}, V.M.~Ghete, N.~H\"{o}rmann, J.~Hrubec, M.~Jeitler\cmsAuthorMark{1}, W.~Kiesenhofer, V.~Kn\"{u}nz, M.~Krammer\cmsAuthorMark{1}, I.~Kr\"{a}tschmer, D.~Liko, I.~Mikulec, D.~Rabady\cmsAuthorMark{2}, B.~Rahbaran, C.~Rohringer, H.~Rohringer, R.~Sch\"{o}fbeck, J.~Strauss, A.~Taurok, W.~Treberer-treberspurg, W.~Waltenberger, C.-E.~Wulz\cmsAuthorMark{1}
% ...
\vskip\cmsinstskip
% ...
1:~~Also at Vienna University of Technology, Vienna, Austria\\
2:~~Also at CERN, European Organization for Nuclear Research, Geneva, Switzerland\\
% ...

\end{sloppypar}
\end{document}